# THE ORIGIN OF COSMIC RAYS:
## HOW THEIR COMPOSITION DEFINES THEIR SOURCES & SITES, AND THE PROCESSES OF THEIR MIXING, INJECTION AND ACCELERATION

RICHARD E. LINGENFELTER
Center for Astrophysics and Space Sciences, University of California San Diego, La Lolla CA 92093 USA;
rlingenfelter@ucsd.edu

> *"any theory of the origin of cosmic rays cannot expect serious success unless it rests on a detailed analysis of the observed composition of primary cosmic radiation."*
> V. L. Ginzburg & S. I. Syrovatskii, *Origin of Cosmic Rays,* 1964
.

## ABSTRACT

Galactic cosmic-ray source compositions, $(Z/H)_{GCRS}$ from H to Pb and $\sim 10^8$-$10^{14}$ eV, differ from solar-local interstellar, $(Z/H)_{SS}$ or $(Z/H)_{ISM}$ by $\sim$20-200x. Both are mostly just mixes of core collapse (CCSN) and thermonuclear (SN Ia) supernova ejecta. The $(Z/H)_{ISM}$ come from steady unbiased accumulation over $\sim$Gyrs. But the cosmic ray mass mixing ratio, universal ISM/CCSN $\sim$4:1 of swept-up ISM and $\sim$10x metallicity ejecta show that $(Z/H)_{GCRS}$ come from basic Sedov-Taylor bulk mixing of homologous, expanding CCSN in their OB cluster self-generated superbubbles, further enriched by highly biased grain-sputtering injection during diffusive shock acceleration (DSA). Moreover, this mixing ratio now reveals that the cosmic rays are primarily accelerated as their evolving reverse shock radius and energy passes through their maxima. Refractories and volatiles, first deposited in fast ejecta and ISM grains in freely expanding ejecta, are simultaneously Coulomb-sputtered $F_{CS}$ by turbulent H and He as suprathermal ions into DSA that carries them to cosmic-ray energies. This bulk mixing selectively increases source mix abundancies $(Z/H)_{SM}/(Z/H)_{SS}$ by $\sim$2-10; and injection by grain condensation and implantation fractions $F_{GC}$, by another $\sim$6, while $Z^{2/3}$-Coulomb grain sputtering enrichments $F_{CS}$ give an added $\sim$4-20. Applying these basic processes of mixing and injection to solar system $(Z/H)_{SS}$ produces grain-injected, source-mix $(Z/H)_{SMGI}$ that match major cosmic ray abundances $(Z/H)_{GCRS}$ to ±35% with no free parameters. Independently confirming grain injection, $(Z/H)_{GCRS}$ shows *no* detectable contribution of Fe from SN Ia, although producing $\sim$1/2 Fe in ISM, but there's also *no* dust in SN Ia remnants, unlike CCSN.

## 1. INTRODUCTION

The quest for the origin of cosmic rays was best charted in a pair of truly prescient papers by Baade & Zwicky (1934a,b). There they first identified a powerful new class of novae, that they called "super-novae," and suggested that these were the gravitational collapse and explosion of massive stars into hypothetical, highly compact remnants, that they dubbed "neutron stars." They also suggested that these powerful supernova explosions were the source of the cosmic rays, and as a test of the idea they further predicted that the cosmic rays would include heavy elements, nuclear charge $Z > 2$.

They were right on all counts. For as observations have shown, such core collapse supernovae (SN II & Ib/c) make up the bulk of Galactic supernovae, producing not only all of the neutron stars, but most of the supernova shock energy, making them the principal source of the Galactic cosmic rays.

Supernovae produce more explosive power than any other known sources in the Galaxy and an order of magnitude more power than that required to generate the Galactic cosmic rays. Observations of supernovae in surrounding galaxies of various types suggest that the supernova rate in our Galaxy is 2.8±0.6 per century (Li et al. 2011; Shivvers et al. 2017). Of these ~81% are CCSN, core collapse of young (~3 to 35 Myr), massive (~ 8 to 120 $M_\odot$) stars, whose winds and explosions blow off most of their mass, leaving only a roughly ~2 $M_\odot$ neutron star remnant, though some simply collapse into black holes without a supernova explosion. The remaining ~19% of the supernovae are the SN Ia explosions of smaller, much older stars that have evolved into white dwarves and are accreting from binary companions, and eventually undergo a thermonuclear explosion blowing off all of their mass and leaving no remnant.

Both types of supernovae release roughly the same $\sim 10^{51}$ ergs in ejecta and shocks (e.g. Nomoto 1984; Woosley & Weaver 1995; Branch & Wheeler 2017) generating a mean explosive power of $\sim 10^{42}$ ergs s$^{-1}$. The cosmic rays with a mean interstellar energy density of $w \sim 10^{-12}$ ergs cm$^{-3}$ (Cummings et al. 2016) filling a Galactic volume, V, and a mean Galactic residence time, $\tau$, require (e.g. Lingenfelter 2013) a sustaining power $Q \sim wV/\tau$, or $\sim wcM/x \sim 10^{41}$ ergs s$^{-1}$, where $x \sim \rho\tau c \sim \tau cM/V$, with the velocity of light, c, the mass of Galactic gas M $\sim 10^{43}$ g, or about 10% of the mass of the Galaxy, and the mean cosmic ray path length, $x \sim 5$ g cm$^{-2}$, as determined by local cosmic ray elemental abundance measurements of the energy dependent secondary/primary ratios, such as Be/CNO and B/CNO, of secondary nuclei produced by spallation of primary nuclei during propagation. Thus the Galactic cosmic rays can be produced by supernova shocks, if they gain ~ 10% of the shock energy.

The most efficient cosmic ray acceleration, as Axford (1981) emphasized, occurs in strong shocks expanding in low density, hot, fully ionized HII regions and supernova remnants can indeed generate cosmic rays with such an efficiency. Strong shocks with compression ratios $s \sim 3$ to 4 can also produce power-law spectra at relativistic energies with an index $\gamma \sim (s + 2)/(s – 1)$ of -2 to -2.5. This range is quite consistent with the cosmic-ray proton spectral indexes of around -2.2, implied by cosmic ray produced, high energy pion-decay gamma rays in both isolated supernova remnants in the interstellar medium (Acero et al. 2016) and those collectively clustered in superbubbles (Ackermann et al. 2011; Aharonian et al. 2018).

Moreover, that source spectral index is in excellent agreement with the values of -2.2 to -2.4, determined from the local equilibrium cosmic-ray index of -2.7 (e.g. Engelmann et al. 1990; Obermeier et al. 2012) with a -0.33 to -0.5 energy-dependent cosmic ray escape lifetime in the Galaxy. That spectral index shift reflects the steepening of the source spectrum during propagation, measured (Obermeier et al. 2012; Adriani et al. 2014; Aguilar et al. 2016) from the spallation production ratio of secondary B to parent CNO nuclei as a function of energy.

The heavy elements in the cosmic rays have now been measured over the energy range from about from $\sim 10^8$ eV to at least $10^{14}$ eV, and perhaps even $\sim 10^{18}$ eV. They all appear to have essentially the same power law energy spectrum of about -2.7, and their relative abundances in the local cosmic rays are found to be effectively independent of energy, as seen in Figure 1.

Detailed elemental and isotopic abundances have now been measured from H to U (Binns et al. 1989; Engelmann et al. 1990; Rauch et al. 2009; Ahn et al. 2010; Donnelly et al. 2012;

Cummings et al. 2016; Murphy et al. 2016). These measurements not only further confirmed Baade and Zwicky's predictions, but revealed broader sources, mixing and injection processes. As Ginzburg and Syrovatskii (1964) foresaw, cosmic ray abundances have provided a wealth of data whose diagnostic value has surpassed all expectation.

The cosmic ray abundance ratios of the metals, Z>2, probe critical nuclear, atomic and even solid state processes in addition to plasma processes, that are not accessible from the study of the H and He ratio alone, which is actually dominated by the interstellar medium unlike the metals. In particular, the metals give multiple measures of the mass mixing ratio of the ejecta and the ambient interstellar medium (ISM) in the cosmic ray acceleration region, together with the nucleosynthetic yields of the ejecta that can identify the type of supernova or other source. They also probe major atomic and mineralogical processes, including differential ionization and grain condensations fractions that determine the relative elemental ion injection factors that dominate both the cosmic ray abundances and their overall acceleration efficiencies. Thus the metals measure major processes not accessible from plasma studies alone.

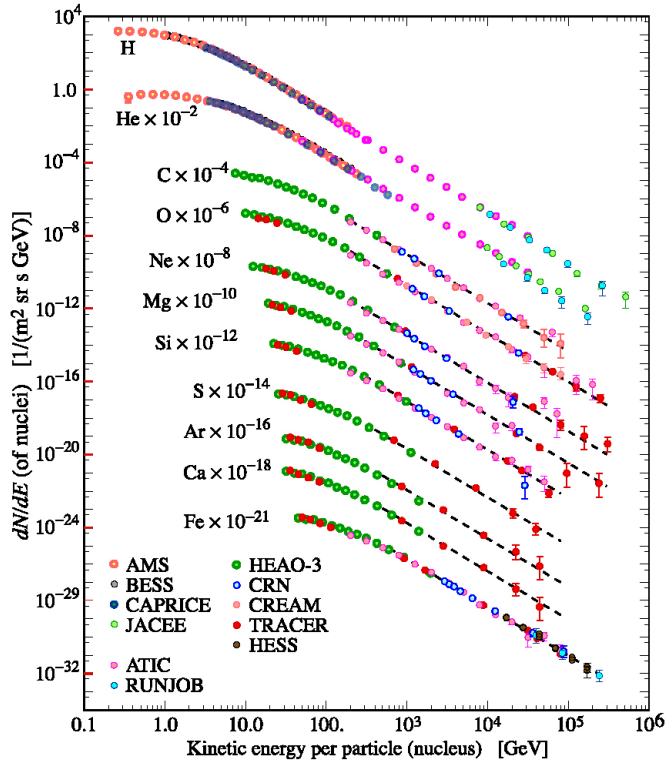

**Figure 1.** The local cosmic ray elemental energy spectra measured by various experiments, showing their essentially constant power law energy spectral index of ~ -2.7, and constant abundance ratios independent of energy (Beringer et al. 2012, and the references therein).

First, we find that these cosmic ray abundances Z/H are not simply solar, or local ISM, but are enriched by factors of ~20 to 200 times solar system values (Lodders, 2003, predominantly C1 chondritic meteorites). Further, we show that these measurements quantitatively define the *driving processes* of supernova source mixing and injection that can determine the cosmic ray elemental abundances to within ±35% with no free parameters. Moreover, we find that the spread of the cosmic-ray bulk mass mixing ratio of swept-up ISM to CCSN ejecta ~$4.3_{-2.0}^{+2.4}$, based on the best-measured cosmic ray source ejecta mass fraction $F_{EJ}$ ~$19_{-6}^{+11}$% (Murphy et al 2016), matches very well the early range of the intrinsic, homologous ratio calculated (e.g. McKee & Truelove 1995; Truelove & McKee 1999, 2000). Basically, this is the growing accumulation and mixing of the fast ejecta and shocked ISM swept-up up by it, in supernova remnants during the period right after the onset of the turbulent, strong shock-generating, Sedov-Taylor stage of supernova expansion and mixing. The need for grain injection further constrains any metal-accelerating remnants to the hot $\sim 10^6$ K, tenuous $\sim 0.001$-$0.01$H cm$^{-3}$ superbubbles, where the grains will not be wholly destroyed by the reverse shock. Altogether these abundance measurements clearly establish the major sources of cosmic-ray injection and acceleration, and also delineate the lesser contributions of thermonuclear SNIa supernovae, Wolf-Rayet winds, asymptotic giant branch stars, and r-process elements potentially in both CCSN and binary neutron star mergers.

Independent of the specific site of cosmic-ray acceleration, there are a number of other processes operating both on the generation of their composition and energy spectrum at their source and on the subsequent modification of these properties to what we now observe. The various elemental and isotopic abundance ratios of the cosmic rays are the most extensive and diverse tracers of all, and place powerful constraints on these processes. In order to determine the cosmic-ray spectrum and composition generated by the cosmic-ray sources, the experimental groups have first corrected their measurements for detector sensitivity and biases, the effects of nuclear spallation in the earth's atmosphere, if needed, and both geomagnetic and heliospheric modulation, to determine the local interstellar cosmic ray abundances, and the Voyager Mission (Cummings et al. 2016) has now directly measured the interstellar cosmic rays. All then model the effects of their propagation and spallation in the interstellar medium enroute from their sources in order to derive the cosmic ray source composition, $(Z/H)_{GCRS}$.

Although various cosmic ray acceleration processes have been proposed, the most likely and most efficient appears to be diffusive shock acceleration (DSA) of suprathermal ions injected into supernova blast waves. This process, developed by Axford et al. (1977), Bell (1978a,b), Blandford and Ostriker (1978), Axford (1981), Jokipii (1982), and Ellison et al. (1997), shows that such ions in the shock-generated turbulent medium both behind and ahead of the forward, as well as the reverse shock can diffuse back and forth through the shock and be *repeatedly* accelerated with a high overall efficiency of ~10% or more.

The similar spectral indices of the various elements in the cosmic ray source and their constant abundance ratios (Figure 1) are also quite consistent with diffusive shock acceleration, which does not differentiate between elements at the same ultra-relativistic rigidity, which is proportional to their kinetic energy divided by their nuclear charge. Thus the observed differences in the elemental source abundance ratios with respect to solar system and interstellar values are not expected to result from plasma acceleration processes (e.g. Ellison et al. 1997; Ohira et al. 2016), except perhaps for small spectral variations if shock injection is significant (Hanusch et al., 2019).

Thus we explore in detail just what the measured cosmic ray elemental abundances and other observations reveal about the nature of their sources and their mixing and injection.

This all develops in the powerful homologous mass mixing of elements in the slowing of the supernova ejecta plasma by the reverse shock of the Sedov-Taylor phase of the expansion once the mass swept up from the surrounding interstellar medium becomes comparable to that of the ejecta. These two components are each divided into gases and solids, --- fast ejecta grains that condensed out in the early free expansion and fast but previously condensed older ISM grains, shock accelerated (Ellison et al., 1997) by the blastwave that swept them up. The fast grains in turn swept up and implanted both heavy refractories and volatiles (e.g. Bibring, et al. 1974; Audouze et al. 1976; Deneault, Clayton & Heger 2003 during the reverse shocks). Simultaneously there is selective injection of these mixed elements into the accelerating shocks as suprathermal ions by sputtering of the condensed and implanted elements in the fast grains, colliding primarily with H and He in the turbulent remnant.

   The fresh supernova ejecta and swept-up, older interstellar material are the two main components of the cosmic ray source mix. The bulk compositions of the ejecta are determined from nucleosynthesis calculations for various stellar progenitor masses and supernova types. Those of the ISM come from optical observations of interstellar gas and dust, and laboratory measurements of meteoritic dust. The relative fractions of refractory and volatile elements that were either condensed or implanted into grains are also particularly important factors. They determine any added abundance enrichment of each element from differential injection into the supernova shocks as suprathermal ions from sputtered refractories and volatiles in fast ejecta grains (Cesarsky & Bibring 1981; Lingenfelter et al. 1998) and in accelerated ISM grains (Epstein 1980; Meyer et al. 1997; Ellison et al. 1997). This is as opposed to the common alternative shock acceleration of the most easily ionized of the highly volatile elements of the ejecta and ISM that remained in the gas phase, especially H, which sets the base line abundance.

   Cosmic ray ejecta mixing is the *key* to cosmic ray acceleration. For it is in fact *the measure* of the shocked supernova remnant mix during the most effective period of such acceleration, and it defines the time frame in which that acceleration occurs in the remnant expansion.

   The basic model of supernova remnant mixing of ejecta and interstellar medium is drawn from the pioneering explosion studies of Taylor (1946) and Sedov (1959). This mixing has long been recognized as a result of the slowing down of freely expanding young supernova remnants after they have swept up a mass of interstellar medium greater than that of the supernova ejecta (e.g. Shklovskii 1962). More recently, McKee & Truelove (1995) and Truelove & McKee (1999) have shown that this mixing is a basic process in the homologous transformation of remnants from free expansion to adiabatic, or Sedov-Taylor expansion, tracing the growth of both forward and reverse shocks, turbulent mixing of swept-up ISM and expanding ejecta, and its increasing ISM/Ejecta mass ratio, before finally sweeping up enough gas to enter the radiative, or "Snow Plow" stage, where it radiates away most of its remaining kinetic energy and becoming subsonic.

   The cosmic ray source ISM/Ejecta mass mixing ratio has also recently been found to be a rather broad universal constant of ~4±2 both in individual elemental cosmic ray abundances and in the Galactic evolution of cosmic-ray spallation products Be and B in old halo stars. As we discuss in detail below (section 5) this range of cosmic ray source mixes strongly suggests that the cosmic rays are in fact primarily accelerated from the shocked ejecta-ISM mix in the remnants within a couple doubling times following the nominal onset of the Sedov-Taylor stage at an ISM/Ejecta ratio of ~1.6 (McKee & Truelove 1995; Truelove & McKee 1999, 2000).

   The earliest measures of supernova mixing and acceleration of fresh ejecta in the cosmic rays have come from extensive observations of old halo star abundances of Be and B, which are produced (Reeves et al. 1970) almost entirely by cosmic ray spallation. The measurements of their evolving enrichment throughout the Galaxy from the earliest times also define the evolving cosmic-ray metallicity and have directly shown (e.g. Tatischeff & Gabici 2018, & references therein) that the cosmic rays *cannot* be accelerated from the interstellar medium *alone*, but must include a major contribution from highly enriched supernova ejecta mixed with swept-up interstellar medium, as will be discussed in detail in Section 2. Comparison (Figure 2) of the solar system composition, *(Z/H)$_{SS}$*, with the Galactic cosmic ray source composition, *(Z/H)$_{GCRS}$*, determined from extensive recent measurements of the local cosmic ray source abundances of the best measured elements, extending all the way up to Pb, clearly shows that they do *not* match with solar system values. Instead, the cosmic rays are greatly enriched by factors of ~20 to ~200 times solar system and local interplanetary medium values. Most individual cosmic ray abundances also differ from one another by factors of as much as ~10 times.

But noting that a subset of the cosmic ray abundance ratios, several of the more refractory elements (Mg, Al, Si, Ca, Fe, Ni, Sr, Zr and Ba) with ratios from 90-160 times solar are roughly the same as one another to within a factor of 2, some have suggested (e.g. Meyer et al. 1997) that these cosmic rays are accelerated almost entirely out of the interstellar medium. However, such similarity is also expected from the CCSN s-process ejecta, since that is the primary source of the interstellar medium (e.g. Timmes et al. 1995). Moreover, the highly refractory, most fully condensed, grain-forming, and most strongly injected elements show the strongest ejecta mixing effect, while all the other less refractory, only partially condensed elements do not. So only these refractories show the ejecta composition similarity with the interstellar medium.

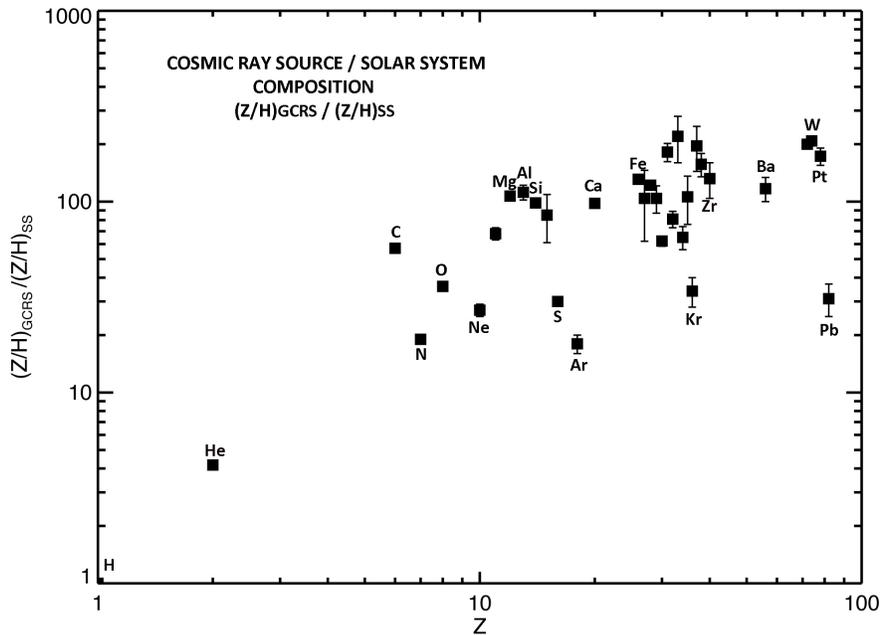

**Figure 2.** Both large general differences and a subset of closer refractory similarities between the cosmic ray source and solar system composition ratios are shown by the ratio $(Z/H)_{GCRS}/(Z/H)_{SS}$ as a function of elemental number $Z$ from the Galactic cosmic ray source abundances relative to H, $(Z/H)_{GCRS}$. The large fundamental differences reflect the basic mixing process of expanding supernova ejecta, while similarities reflect their major common source in CCSN ejecta. $(Z/H)_{GCRS}$ are determined from local cosmic ray composition measurements (Engelmann et al. 1990/Cummings et al. 2016, $1 < Z < 28$; Rauch et al. 2009/Murphy et al. 2016, $26 < Z < 40$; Binns et al. 1989, $40 < Z < 70$; Donnelly et al. 2012, $Z > 70$) and $(Z/H)_{SS}$ solar system values (Lodders, 2003).

Still both these large differences and selected similarities between the cosmic ray source and the solar system compositions offer clear clues as to their causes. As noted, detailed analyses of these source abundances relative to those of the interstellar medium show that these differences can be almost entirely resolved without any free parameters by assuming that the source abundances seen by the accelerating supernova shocks result from just the two processes, bulk mixing of ejecta with the ISM and selective injection of these mixed elements into the accelerating shocks by grain sputtering.

In Sections 2 and 3 we apply the two basic mixing and sputtering processes sequentially to solar system elemental abundances (Lodders 2003) to show, as Rauch et al. (2009) and Murphy et al. (2016) have done with the isotopic abundances, just how remarkably revealing a comparison of the cosmic ray source abundances with the mix of the supernova ejecta and the interstellar medium can be. We see particularly how it organizes those abundance ratios to clearly define the injection process in terms of the fraction of each refractory element that condensed into, and volatile element that was implanted into high velocity grains, and was then sputtered off as a suprathermal ion. Taking that procedure further, we find that next applying the elemental charge $Z^{2/3}$-dependent Coulomb sputtering cross section or relative enrichment and then the grain condensation fraction, moves the fit of the abundances closer to and finally into agreement with the cosmic ray observations. Thus the procedure demonstrates both the need for such processes and their effectiveness, and clearly reveals the remaining process.

In Section 4, using the ability of these processes to explain the bulk of the cosmic ray metal abundances primarily in terms of core collapse supernova nucleosynthesis, we define the contribution of thermonuclear SN Ia supernovae, Wolf-Rayet winds, asymptotic giant branch stars, and the relative production of r-process elements by core collapse supernovae and neutron star merger kilonovae.

Finally, in Section 5, we explore the major implications of these processes in the context of the other constraints on the supernova sources and sites of cosmic ray source mass mixing, grain injection and diffusive shock acceleration. Most importantly, we find that this uniform Galactic range of cosmic-ray bulk mass mixing, ISM/CCSN $\sim 4.3^{+2.4}_{-2.0}$ (Murphy et al. 2016), matches that of the fundamental, homologous range of that mixing ratio calculated (McKee & Truelove 1995; Truelove & McKee 1999; 2000) in evolving young supernova remnants at the onset of turbulent, shock-generating Sedov-Taylor expansion, outlining the interconnected spatial and temporal structures of the cosmic-ray source mixing, injection and acceleration.

Moreover, the occurrence of strong $Z$-dependent enrichment of cosmic rays from sputtering interactions between fast supernova grains and swept-up gas also requires that these grains not be >90% destroyed by grain-grain collisions in reverse shocks, which is expected if supernovae are expanding in interstellar gas denser than n > 0.1 H cm$^{-3}$ (Bianchi and Schneider 2007) and unlike ion injection by only limited, but not complete, sputtering does *not* result in any Z-dependent enrichment.

## 2. COSMIC RAY SOURCE MIXING

We now consider the evidence for mixing of heavy elements in the supernova ejecta and swept-up interstellar medium in the cosmic ray source material, the fundamental nature of the supernova remnant mixing *by mass*, and finally what a comparison of the cosmic ray source composition with that source mix reveals about the additional preferential injection of these elements as suprathermal ions sputtered from fast refractory grains into the accelerating shocks. As observations have shown, nearly all of the refractory elements condense into fast dust grains very early in the freely expanding ejecta, as it rapidly cooled to as low as 20 K in the moving frame. With the transition to Sedov-Taylor expansion came massive ejecta-ISM mixing and heating of the grains back up to ~ 1000 K (Bianchi & Schneider 2007) and their partial sublimation in the reverse shock. Those grains that survived destruction in that shock were decoupled from the slowing, shocked plasma and moved on ahead at suprathermal velocities close to their expansion velocity of ~3,000 km s$^{-1}$, or energies of several MeV per atom for the major refractory elements, into the much slower, swept up, compressed and mixing ISM. There, in interactions with turbulent H, He and other volatiles, they were sputtered off as suprathermal ions at similar velocities, or ~10 times 150 km s$^{-1}$ that of the 10$^6$ K H gas, injecting the metals into the diffusive shock acceleration process to be selectively carried to cosmic ray energies.

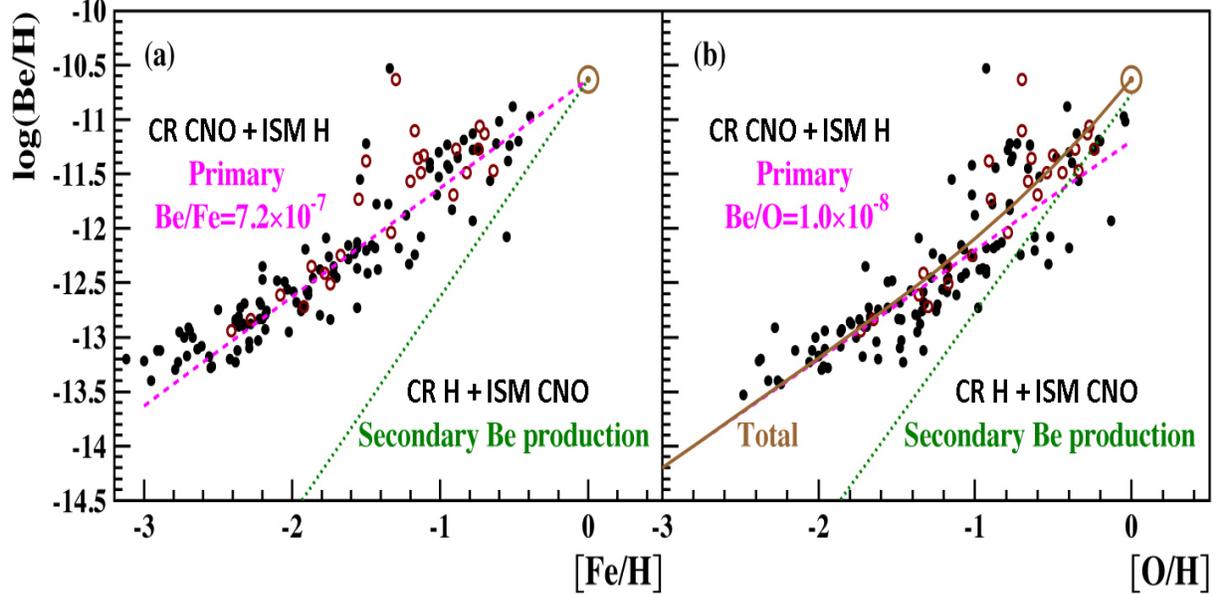

**Figure 3.** The Be/H in old halo stars show that the Be is directly proportional to the primary supernova production rate of Fe (a) and O (b) as expected from spallation of supernova ejecta CNO in the cosmic rays, and not to their secondary interstellar accumulation rate as expected from spallation of CNO in the interstellar medium (adapted from Tatischeff & Gabici 2018).

The earliest and most far reaching evidence for accelerated supernova ejecta in the cosmic rays, however, came not from measurements of the various current, local cosmic ray abundances, but instead from quite different studies of the long-term evolution of the cosmic ray abundances of just the CNO group, determined from abundances of Be and B in the atmospheres of old halo stars over the last ~10 Gyr. These two exceedingly rare elements, $(Be/H)_{SS}$ ~ 3x10$^{-11}$ and $(B/H)_{SS}$ ~ 7x10$^{-10}$ (Lodders, 2003), are unique in that they cannot be produced by standard stellar nucleosynthesis and, as Reeves, Fowler and Hoyle (1970) first suggested, they are produced instead by spallation of CNO in cosmic ray interactions in the interstellar medium. They are produced by two reciprocal processes, the spallation of interstellar CNO atoms primarily by cosmic ray protons and helium and inversely by spallation of cosmic ray CNO nuclei colliding primarily with interstellar H and He atoms. Currently the contribution of spallation of interstellar CNO nuclei is about twice that of cosmic ray nuclei (Ramaty et al. 1997).

Measurements (Gilmore et al. 1992; Duncan et al. 1992; and others, see review by Tatischeff & Gabici 2018) of the Be and B in old halo stars born in the early Galaxy with low interstellar gas metallicity, however, showed that at that time the Be and B resulted just from spallation of cosmic ray CNO. For these Be and B abundance ratios were constant relative to both O and Fe, and within a factor of ~2 of the present values, while the interstellar $(Fe/H)_{ISM}$ grew by orders of magnitude from 10$^{-3}$ to 10$^{-1}$ $(Fe/H)_{SS}$. Thus the Be and B production was proportional to the Galactic supernova rate and ejecta production, and had to be produced by spallation of some fraction of their ejecta in the cosmic rays, since the elemental yields of the dominant core collapse supernovae are independent of their initial metallicity in this range (e.g. Woosley & Heger 2007). There is no significant measureable evidence of any component of the Be and B production that was proportional to the interstellar metallicity, which varied by a factor of $>10^2$, as would have been expected if they were produced solely by cosmic ray spallation of the interstellar medium (Duncan et al. 1992, 1997; Ramaty et al. 1997, 2000; Alibes et al. 2002).

The estimated fraction by mass, $F_{EJ}$, of the highly enriched supernova ejecta in a mix with the very low metallicity interstellar medium in the cosmic ray source material, necessary to satisfy the Be and B observations, was at least 15% (Ramaty et al. 1997), or a bulk cosmic-ray mass mixing fraction of swept-up ISM/SN ejecta <6. Subsequent calculations (Alibes et al. 2002) of the evolution of the spallation-produced Be and B versus Fe/H, as a function of the supernova ejecta mass fraction in the cosmic ray source, compared to Be and B abundances measured in halo stars over the range of Galactic metallicities, $(Fe/H)_{ISM}$ from $10^{-3}$ to 1 $(Fe/H)_{SS}$, gave a best fit supernova ejecta mass fraction $F_{EJ}$ ~25 ±15%, or a mass mixing fraction ~3. But this ejecta fraction is *not* the elemental abundance ratio of the source mix, which is dominated by the highly enriched supernova ejecta. With such an ejecta mass fraction this mixing increases the elemental cosmic ray source abundances by factors of ~ 2 to ~10 times those of the interstellar medium.

Moreover, these studies show that this supernova ejecta mass fraction in the cosmic ray source mix, which essentially describes the initial mixing of the ejecta with the interstellar medium, is not just a local or transient value, but a global property of cosmic ray source abundances, since these stars sample the whole Galaxy for over ~ 10 Gyr.

Although alternative models for Be and B production without cosmic ray acceleration of supernova ejecta have been explored, none are consistent with other constraints (e.g. Tatischeff & Gabici 2018 and references therein) and all fall short by more than an order of magnitude. Thus, the observed Galactic evolution of Be and B abundances requires that a constant fraction of the supernova ejecta CNO be accelerated to cosmic ray energies. The simplest acceleration process would be for the ejecta of a supernova to be accelerated by its own shocks (Lingenfelter et al. 1998). This is also quite consistent with what is expected (McKee & Truelove 1995; Truelove & McKee 1999) in homologous, expanding supernova remnants at the onset of the Sedov-Taylor stage of expansion (Section 5.2), and a core feature of cosmic ray acceleration that easily explains its universality.

A constraint on such acceleration, however, is set by ACE/CRIS limits (Wiedenbeck et al. 1999) on the abundance of cosmic ray radioactive $^{59}$Ni, which decays to $^{59}$Co with a half-life of $7.6 \times 10^4$ yr only by bound electron capture, and cannot significantly decay once the $^{59}$Ni is accelerated and its electrons are stripped off (Casse & Soutoul 1975; Soutoul et al. 1978). The standard non-rotating star nucleosynthesis models of CCSNe (Woosley & Weaver 1995, Table 5A) predict that radioactive Ni makes up ~60% of the combined freshly produced mass 59 Co and Ni nuclei, or $^{59}$Ni/$^{59}$Co ~1.5, although recent models (Limongi & Chieffi 2018, Table 30) of both rotating and non-rotating stars predict only 37 to 35%, or $^{59}$Ni/$^{59}$Co ~0.56, instead.

Thus, since the measured (Wiedenbeck et al. 1999) cosmic ray abundance of stable $^{59}$Co/$^{60}$Ni nuclei is ~0.182 ±0.021, we would expect the fresh undecayed cosmic ray $^{59}$Ni /$^{60}$Ni to lie between 0.56 and 1.5 times that, or ~(0.10 and 0.27) ±0.021. However, the measured cosmic ray $^{59}$Ni /$^{60}$Ni is <0.055, which is consistent with the 0.049±0.012 abundance that might be expected purely from secondary production by spallation of heavier cosmic rays during subsequent propagation with no significant evidence of $^{59}$Ni remaining. If supernova ejecta alone were being accelerated to cosmic ray energies, they could not have been so in $<10^5$ yr.

But as the Galactic Be and B evolution alone shows, that cannot be the case. The mass mixing of old $>>10^5$ yr, well-decayed, swept-up ISM into the fresh CCSN ejecta greatly cut the relative abundance of $^{59}$Ni in the remnant itself and its significance in the cosmic rays. Mixed with a mean $F_{EJ}$ ~20% of ejecta $^{59}$Ni plus 80% of none in the ISM, the expected $^{59}$Ni / $^{60}$Ni ~ (0.10 to 0.27) ±0.021 is cut by 5 to just (0.020 to 0.054)±0.021. But that is offset by the factor of ~10 Ni metallicity of the ejecta, for a net cut of a factor of 0.6 to (0.06 to 0.16)±0.021. However, with most of the 10 to 20 $M_\odot$ CCSN remnants clustered in low density ~0.001H cm$^{-3}$ superbubbles around OB associations, the cosmic-ray accelerating phase of the Sedov-Taylor expansion is stretched from relatively short <$10^4$ yr scales observed in isolated radio and high energy gamma ray remnants in the typical ~1H cm$^{-3}$ ISM to as much as the $^{59}$Ni decay half-life. There radioactive decay by bound electron capture can also significantly cut the expected $^{59}$Ni /$^{60}$Ni by roughly an additional 1/2 to only ~(0.03 to 0.08) ±0.021, less than 2-σ statistical errors. Thus, the combined effects of decay, mixing dilution and rotation yields can now account for the absence of $^{59}$Ni in cosmic rays.

The general mixing in the cosmic ray source of fresh, high-metallicity supernova ejecta with the swept-up interstellar gas and dust has been better quantified by a number of different measurements of local cosmic ray elemental and isotopic ratios. These studies began (Higdon & Lingenfelter 2003) with the measured (Binns et al. 2001) cosmic ray $^{22}$Ne/ $^{20}$Ne ratio of 0.366±0.015, which is fully 5.0±0.2 times that of the solar system ratio. Using the core collapse supernova s–process yields (Woosley & Weaver 1995) and allowing for large uncertainties in the added yields of these isotopes in Wolf-Rayet winds (Maeder & Meynet 2000), this isotope ratio implies a mass fraction, $F_{EJ}$ of 18±5%. Similar mass fractions of about 20% have since been found (Binns et al. 2005, 2006, 2007), using later Wolf-Rayet models, for the $^{22}$Ne/$^{20}$Ne, and the other three largest cosmic ray isotopic deviations from solar system values, $^{12}$C/$^{16}$O, $^{58}$Fe/$^{56}$Fe, and <3 Myr-old radioactive $^{60}$Fe/$^{56}$Fe from CCSN (Binns et al., 2016, Fig.3), in addition to the $^{59}$Ni (Wiedenbeck et al. 1999; Binns et al. 2008b).

Calculations (Lingenfelter et al. 2003) of cosmic ray elemental enrichment ratios of the ThU/Pt group from supernovae also suggests a similar ejecta mixing ratio $F_{EJ}$ close to 20% for the products of explosive r-process supernova nucleosynthesis. These calculations based on Kratz et al. (1993) of the radioactive Actinide/Pt group abundance ratio predicted a cosmic ray source value of 0.027±0.005, which is quite consistent with the latest balloon measurements (Combet et al. 2005; Donnelly et al. 2012) of ~0.032.

Subsequent studies (Lingenfelter & Higdon 2007; Binns et al. 2007, 2008a,b) of the origin of the cosmic ray composition explored not only the CCSN / ISM mixing ratio, but also the grain condensation and sputtering enrichment. Recent higher energy measurements (Ahn et al. 2010) of the abundance ratios of cosmic ray He, C, N, Mg, Si and Fe relative to O up to ~4 TeV/nucleon also found a good fit with a 20% ejecta mixing ratio.

In the most extensive study, including new cosmic ray isotopic abundances measurements of heavier elements of Zn to Zr on the Antarctic long-duration balloon flights *TIGER* and *Super-TIGER,* Rauch et al. (2009) and Murphy et al. (2016) have found best-fit mixing mass fractions of around 19% newly synthesized, high-metallicity massive star ejecta and 81% old interstellar gas and dust. Moreover, comparing the cosmic ray source composition with this source mix, they also quantified the subsequent elemental injection biases that further enrich the cosmic ray composition, as did Ahn et al. (2010), which we discuss further in Section 3.

Thus, the local cosmic ray measurements all suggest an essentially constant cosmic ray source component of $F_{EJ}$ ~20% high metallicity ejecta of newly synthesized nuclei over a wide energy range of at least 4 decades of energy from ~100 MeV/nucleon up to ~4 TeV/nucleon, for the nominal ~ 20 Myr lifetime of the current local < 1 kpc cosmic rays. As we also saw, analyses of old halo star abundance ratios of Be, produced primarily by the spallation of cosmic ray CNO, give a similar constant ratio of ~25% in the far older and much more distant cosmic ray position extending out to ~ 10 kpc and back over >10 Gyr, when the average Fe metallicity of the interstellar medium was as low as $(Fe/H)_{ISM} \sim 10^{-3} (Fe/H)_{\odot}$.

All of these measurements show that the individual elemental abundance ratios in the core collapse supernova source mix can be simply defined by a single mixing mass fraction, $F_{EJ}$, of highly enriched supernova ejecta mixed with a swept-up interstellar mass fraction, $F_{ISM} = 1 - F_{EJ}$, or a bulk ISM/CCSN mass ratio of $\sim(1 - F_{EJ})/F_{EJ}$. The resulting elemental abundance ratios of the supernova source mix relative to solar system abundances are

$$(Z/H)_{SM/SS} = F_{EJ} (Z/H)_{EJ/SS} + F_{ISM} (Z/H)_{ISM/SS} = F_{EJ} (Z/H)_{EJ/SS} + (1 - F_{EJ})(Z/H)_{ISM/SS}. \quad (1)$$

The source mix abundances, resulting from this ubiquitous supernova ejecta mass fraction $F_{EJ}$ of 20% mixed with an interstellar medium mass fraction $F_{ISM}$ of 80%, or a bulk ISM/CCSN mass mixing ratio of ~4, are very different from those of the solar system (Lodders 2003). Since the present local interstellar medium $(Z/H)_{ISM} \sim 1.32 (Z/H)_{SS}$ (Timmes et al. 1995), the source mix reduces to simple ratios of ejecta abundances to solar system values, $(Z/H)_{SM}/(Z/H)_{SS} \sim 0.2$ $(Z/H)_{EJ}/(Z/H)_{SS} + 1.06$, which gives individual abundance ratios varying by a factor of 10, as can be seen in Figure 4. Even though the swept-up ISM dominates the source mix mass by a factor of ~4, the freshly synthesized, high metallicity ejecta, still greatly dominates the elemental source mix abundances $(Z/H)_{SM}$ by factors of ~2 to 6.

The individual elemental abundances, $(Z/H)_{EJ/SS}$, of the supernova ejecta combined with the pre-supernova winds are the Salpeter IMF weighted yields calculated, relative to solar system values for the dominant core collapse supernovae, SN II & Ib/c, that make up the bulk (81%, Li et al. 2011) of Galactic supernovae. These yields for both the standard non-rotating star (e.g. Woosley & Weaver 1995; Woosley & Heger 2007) and recent rotating and non-rotating star (Limongi & Chieffi 2018) models of the slow neutron capture s-process nucleosynthesis are the primary process in main sequence stellar evolution. Both of these supernova yields, based on an extensive network of nuclear reactions, have been shown (Timmes et al. 1995; Prantzos et al. 2018) to be consistent with Galactic chemical evolution.

For the rotating star models, Limongi & Chieffi (2018, Table 30) calculated the CCSN elemental nucleosynthetic yields for three fiducial surface rotation velocities, v = 0, 150 and 300 km s$^{-1}$, and defined an Initial Distribution of Rotational Velocities (IDROV), paralleling the Initial Mass Function (IMF). The relative contributions of these three fiducials of the velocity distribution were then calibrated (Limongi & Chieffi 2018; Prantzos et al. 2018, Fig. 4) by the measured Galactic chemical evolution as a function of Galactic metallicity [Fe/H], shifting from a relative weight of 75% for v=300 km s$^{-1}$ yields and 25% for v=150 km s$^{-1}$ from early times at metallicity -3, to 67% v=0 and 33% v=150 km s$^{-1}$ at solar metallicity, 0. Using the latter weights for their fiducial yields, we calculated a current rotating star CCSN source mix abundances shown in Figure 4b, and for the standard non-rotating star model yields (Woosley & Heger, 2007, Fig. 7) we calculated the non-rotating CCSN source mix abundances shown in Figure 4a.

These CCSN abundances for the two models, 4a and b, appear to be quite similar and robust for the major elements measured in the cosmic rays. The elemental differences between these models are generally much less than a 1/2, while their CCSN supernova mix abundance ratios relative to solar system and interstellar values each differed much more widely by ~2 to ~8X.

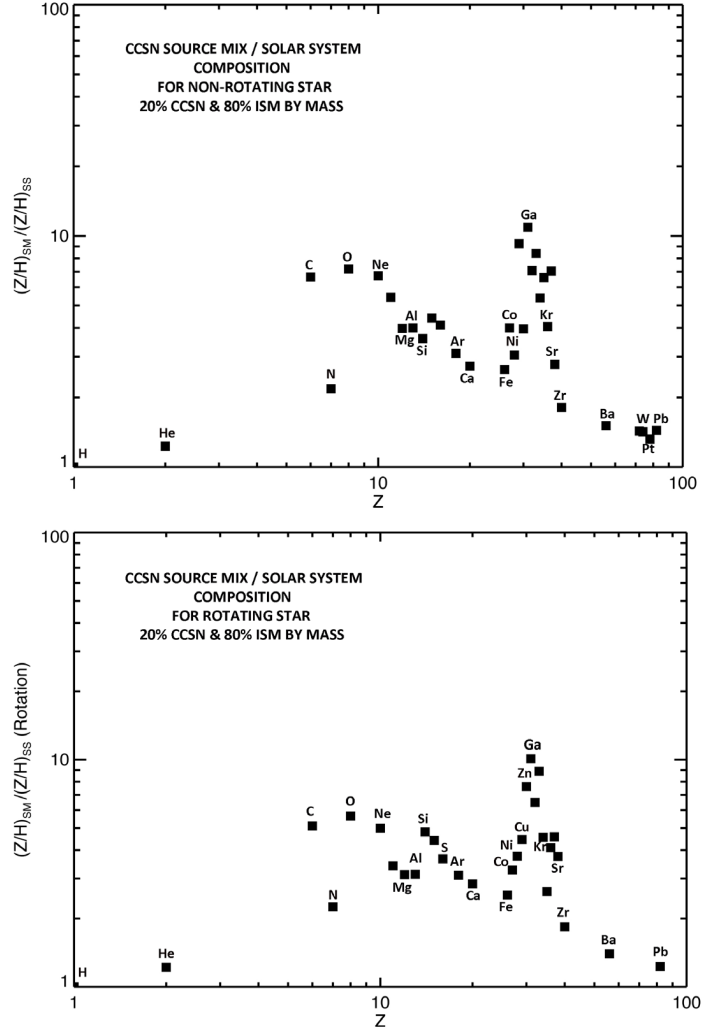

**Figure 4a,b.** The ratio *(Z/H)$_{SM}$/(Z/H)$_{SS}$* of the major individual elemental cosmic ray source mix abundances relative to H, which represents the homologous Sedov-Taylor CCSN mix during the peak period of cosmic ray acceleration, are very different from the solar system, and local interstellar medium values (Lodders 2003). Even though their source mass mix is weighted toward ISM with ISM/CCSN ~ 4:1, or $F_{ISM}$ of 80% swept-up old ISM and $F_{EJ}$ of 20% fresh CCSN ejecta, because the ejecta has a very high metallicity, ~10 times solar, the supernova ejecta still dominates the elemental source mix abundances by factors of ~2 to 6. These supernova ejecta yields, consistent with Galactic chemical evolution, are calculated (a) for the CCSN standard non-rotating star model yield (Woosley & Heger 2007, Fig. 4), and (b) for a weighted mix of 2/3 nonrotating (v=0) and 1/3 rotating (v=150 km s$^{-1}$) star model yields (Limongi & Chieffi 2018, Table 30).

Even the CCSN supernova ejecta abundances, which are dominated ~70% by the more numerous and productive SN II progenitors below ~ 25 M$_\odot$, are only very weakly affected by whether the more massive stars above ~25 M$_\odot$, are able to explode as supernovae contributing to the mix, or whether they fail to explode and collapse in a black hole, taking most of their nucleosynthetic products with them. Various studies (Heger et al. 2003; Smartt 2015; Sukhbold et al. 2016; Mirabel 2017) suggest such masses for the onset for stellar black hole formation, and calculations (Woosley & Heger 2007, Fig. 6) have shown that even if these more massive stars did explode as CCSNe they would make "little difference except for the iron group" and even there they would only increase Fe/Si by 10% to 20%. The other massive stars up to ~120 $M_\odot$ are thought (Woosley & Heger 2007; Branch & Wheeler 2017) to be stripped of H envelopes by mass transfer to binary companions or blown off in stellar winds to become He-rich stars of <6 $M_\odot$ and SN Ib or Ic, accounting for ~20 and 10% respectively of CCSN (Shivvers et al 2017).

The s-process yields, together with important, but limited, contributions to C and N from Asymptotic Giant Branch (AGB) stars and Wolf-Rayet winds, and to the Fe peak from thermonuclear SN Ia supernovae of accreting white dwarfs (e.g. Nomoto et al. 1984, model W7), can account for the solar elemental abundances (e.g. Anders & Grevesse 1989; Lodders 2003) of the bulk of the elements up through Zn at Z = 30. For heavier nuclei all the way up through the actinides, the fast neutron capture r-process of explosive nucleosynthesis (e.g. Kappeler et al. 1989, 2011) competes with the core collapse s-process with each contributing about half overall, but their relative contributions vary from isotope to isotope. The core collapse supernovae are a significant source of the r-process nucleosynthesis, but the newly detected neutron star mergers are also thought to be important sources, and the cosmic ray abundances can provide a measure of their relative strength.

The major differences between the heavy elemental cosmic ray mix and the interstellar medium is that the latter is essentially the unbiased mix of core collapse SN II/Ibc and thermonuclear SN Ia, while the cosmic ray source mix is a highly biased one. This mix, which is dominated by the enriched core collapse supernova ejecta, and the refractory element grain condensation in it, plus those in the swept up interstellar medium, is even further enriched by grain sputtering injection into the accelerating supernova shocks. On the other hand, the cosmic ray source abundances also show that only a negligible fraction of the cosmic ray metals comes from the very highly enriched ejecta of SN Ia supernovae, as we discuss in detail in Section 4.

First, however, we explore what the cosmic ray source abundances reveal about refractory supernova grains and their sputtered suprathermal ion injection in core collapse supernovae. Here, we present a new analysis in Figure 5, comparing the most recent elemental cosmic ray source abundances, $(Z/H)_{GCRS}$, from H to Pb with those of the best-fit source mix, $(Z/H)_{SM}$.

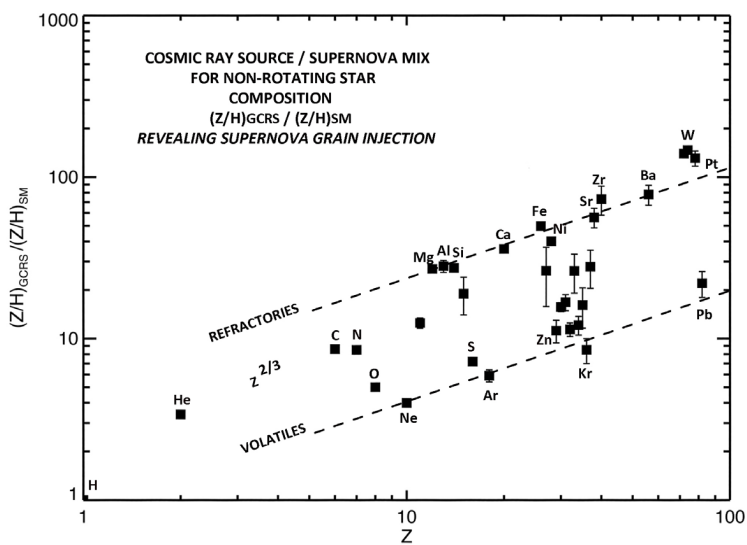

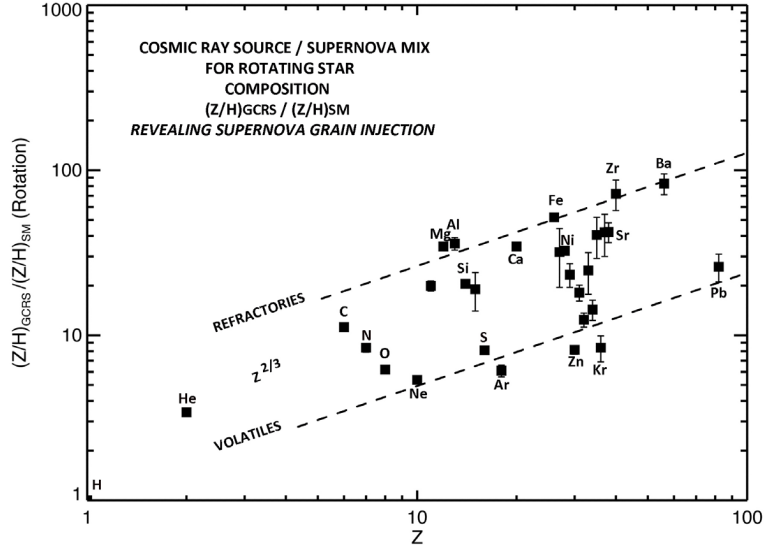

**Figure 5a,b**. Not surprisingly the ratio $(Z/H)_{GCRS}/(Z/H)_{SM}$ of cosmic ray source abundances relative to the seemingly disorganized, but universal 4:1 ISM/CCSN mix, sharply organizes the abundance variations and the injection factors, showing both the correlation of the abundances with a nominal $Z^{2/3}$ (dashed lines) grain Coulomb sputtering factor, $F_{CS}$, and an injection bias constant for refractory grain forming elements versus volatile elements, $C_{R/V}$. The $(Z/H)_{GCRS}$, are calculated from the measured (Engelmann et al. 1990 / Cummings et al. 2016, $1 < Z < 28$; Rauch et al. 2009, $26 < Z < 38$ / Murphy et al. 2016, $30 < Z < 40$; Binns et al. 1989, $40 < Z < 70$; Donnelly et al. 2012, $Z > 70$) local interstellar cosmic ray flux spectra between ~GeV and ~TeV/nucleon, divided by their corresponding abundances in the source mix $(Z/H)_{SM}$ (Figure 4) for the best-fit ~20% ejecta mass fraction of both non-rotating core collapse supernova (Woosley & Heger 2007) and rotating (Limongi & Chieffi 2018).

## 3. COSMIC RAY SOURCE INJECTION

Hints of a grain injection were first recognized (Reeves 1975) from the fact that the relative cosmic ray source abundances of refractory elements, which are the first to condense into grains in a cooling medium, depleting the gas, and the last to sublime in a warming or hot medium, were rather uniformly larger than those of volatiles. Such fast grains are condensed in the rapidly expanding supernova ejecta (Cesarsky & Bibring 1981; Lingenfelter et al. 1998), but they can also be accelerated by supernova shocks out of the older, swept-up interstellar medium (Epstein 1980; Meyer et al. 1997).

A grain connection was further supported (Lingenfelter & Higdon 2007) by measurements (Engelmann et al. 1990) of the cosmic ray source abundances, most clearly of the volatiles, showing an atomic charge $Z$ dependence close to that of the grain sputtering Coulomb cross section $Z^{2/3}$ (Sigmund 1969, 1981; Draine & Salpeter 1979; Tielens et al. 1994). This sputtering also appears as an $A^{2/3}$-dependence for relatively low mass elements $A < 40$, where the elemental mass $A$ is simply equal to $2Z$.

Such sputtering of fast grains can inject both the refractory elements originally condensed in the grains and also the refractories and volatiles implanted in them (e.g. Bibring, et al. 1974; Audouze et al. 1976; Deneault, Clayton & Heger 2003), as suprathermal ions. Thus, as the mean cosmic-ray bulk mass mixing ratio of swept-up ISM to CCSN ejecta ~$4.3_{-2.0}^{+2.4}$ (Murphy et al 2016) indicates, both the peak shock acceleration of the cosmic rays and the $Z^{2/3}$-dependent sputtering enrichment of the suprathermal ions that feed it, occur in the turbulent region of the remnant behind both the forward and reverse shocks during the period right after the onset of the turbulent, strong shock-generating, Sedov-Taylor stage of expansion (McKee & Truelove 1995; Truelove & McKee 1999).

This basic elemental Z-dependent Coulomb sputtering $F_{CS}$ from interactions between two nuclei of charges $z$ and $Z$ is the sputtering cross section $zZ / (z^{2/3} + Z^{2/3})^{1/2}$ (Sigmund 1969, 1981). Since fast grain element interactions are predominantly with ambient H, the enrichment factor for refractories, e.g. $Z > 6$, reduces to $F_{CS} \sim Z / (1 + Z^{2/3})^{1/2}$. Although the enrichment for interactions with He is roughly a constant ~ 2 times that of H, the relative abundance of He/H in the swept-up medium is only ~ 0.1, so that He contributes only about 1/5 as much as H, and only about 1/6 of their total with virtually the same $Z$ dependence. The He contribution becomes just a small addition to the general normalization constant.

Thus the calculated ratio of Coulomb sputtering cross sections between any two cosmic ray elements would be expected to define their relative H sputtering enrichment with respect to their bulk mixing abundances solely as a function of their different charges $Z$, if their sputtered elements, Z and H, are directly injected as suprathermal ions into the accelerating shocks. So we shouldn't be surprised to see in Figure 5 that the measured ratios $(Z/H)_{GCRS}/(Z/H)_{SM}$ of the cosmic ray source abundances divided simply by the uniform 4:1 bulk mass mix from Figure 4 match very well the nominal $Z^{2/3}$ enrichment ratios (dashed lines -- -- --) expected (Sigmund 1969, 1980) from sputtering enrichment by near MeV/atom interactions injecting sputtered suprathermal ions into cosmic ray accelerating shocks. This is also confirmed in the ratio $(Z/H)_{GCRS} / (Z/H)_{SMCSI}$ and the maximum refractory grain enrichment constant relative to volatiles (Figure 6).

These cosmic ray injection ratios, however, are quite distinct from the standard integral "sputtering yield" (Sigmund 1969, 1980) that includes, and is dominated by, a far larger cascade of much lower energy sputtering interactions down to just eV/atom binding energy thresholds. These also result from some initial interactions. But only a very small fraction of these can pass the high energy interaction threshold, required to produce suprathermal ions that seed cosmic ray acceleration. Moreover, the relative abundances of all these with low energy thresholds are shifted by factors of about ~1 to ~10 in their binding energies in addition to their Coulomb sputtering ratios. These standard yields, also match very well with laboratory measurements (e.g. Andersen & Bay, 1980) and supernova grain survival calculations (e.g. Micelotta et al. 2016).

Although a seemingly attractive alternative to the refractory versus volatile bias was preferential acceleration of elements with the lowest first ionization potential (Casse & Goret 1978), which would be most easily and highly ionized in the warm neutral, or partially ionized gas and thus the most easily accelerated. Indeed such an effect was clearly observed in energetic solar flare particles. But, as Axford (1981) first pointed out, such injection would not operate in the hot, fully ionized plasma, where shock acceleration is most efficient, and instead would limit the acceleration to cooler, denser gas where it is least efficient, because of much higher competing ionization losses, rapid Alfven wave damping, and radiative losses. Since many of the most favored elements were the same, it was at first hard to discriminate between the two alternatives observationally. Meyer et al. (1997), however, found several elements where the two differed significantly, and all favored a grain bias. They also noted a strong mass-dependent, $A^{0.8\pm0.2}$, enrichment of the highly volatile cosmic ray elements, N, Ne, S, Ar, Se & Kr, which as later pointed out (Lingenfelter & Higdon 2007) is also quite consistent with the classical $Z^{2/3}$ enrichment expected from grain sputtering injection (Sigmund 1969, 1981) of cosmic rays.

Both the ejecta mixing ratio and the grain connections have now been greatly strengthened by new cosmic ray source composition measurements and analyses of Murphy et al. (2016) from their *SuperTIGER* and the earlier *TIGER* (Rauch et a. 2009). They not only found best-fit mixing mass fractions of around 19% fresh CCSN ejecta and 81% old interstellar gas and dust, but on comparing the cosmic ray source composition with this source mix, rather than that of the solar system, they were able to quantify the elemental injection biases.

They found that those more refractory cosmic ray elements Mg, Al, Si, Ca, Fe, Ni, Sr, & Zr were all roughly ~ 4 to ~4.5 times more abundant with respect to solar system values than the most volatile ones, Ne, S, Ar, Cu, Ge, Se, & Kr. They further found that the best-fit atomic mass dependent enrichment of the highly volatile cosmic ray elements was $A^{0.632\pm0.119}$, which is quite consistent with the ~ $A^{2/3}$ expected from the grain Coulomb sputtering and scattering cross section, for $A/Z = 2$. However, for the more abundant highly refractory cosmic ray elements with tighter uncertainties the best-fit value was just $A^{0.583\pm0.072}$, so the authors concluded only that both the refractory and volatile elements show a mass-dependent enrichment with similar slope.

But $A^{2/3}$ is just approximately the $A$-dependence for the $Z^{2/3}$ Coulomb sputtering dependence, since $A/Z = 2$ only for the limited range of refractories from $20 < A < 40$, while for all of the neutron-rich elements above that, $A$ grows relative to $Z$ with what can be approximated as a power-law dependence, where $Z \sim A/2A^y \sim 0.5A^{(1-y)}$. Thus the $A$-dependent Coulomb sputtering and scattering yield enrichment of such heavy elements that is equivalent to $Z^{2/3}$ is ~ $A^{2(1-y)/3}$.

For the cosmic ray source abundances analyzed by Murphy et al. (2016) the refractories ranged between $^{24}$Mg and $^{91}$Zr, or a relative atomic mass range ($A_{MAX}/A_{MIN}$) of 91/24 = 3.79, with $A/Z$ growing from 2.0 to a maximum of 2.28. This gives a nominal exponent $y = \log (2.28/2) / \log (91/24) = 0.10$, and the expected $Z^{2/3}$ equivalent $A$-dependent index $2(1-y)/3 = 0.60$. That value is quite consistent within 0.24 σ of the best-fit value of 0.583±0.072 (Murphy et al. 2016). Similarly, the volatiles, which ranged from $^{20}$Ne to $^{80}$Br with $A/Z$ rising from 2.0 to 2.32, give a nominal exponent $y = \log (2.32/2) / \log (80/20) = 0.11$, and the expected $Z^{2/3}$ equivalent $A$-dependent index $2(1-y)/3 = 0.59$, and that too is in very good agreement with the best-fit value of 0.632±0.119. Thus, the cosmic ray source abundance analyses of Murphy et al. (2016) would actually provide strong evidence for a $Z^{2/3}$ fast grain sputtering injection of the cosmic ray source mix of core collapse ejecta (Woosley & Heger 2007).

### 3.1 Grain Condensation Fraction

The formation of fast refractory dust grains has been observed in the early, freely expanding ejecta of core collapse supernovae, and, as these high velocity grains are subsequently mixed through ISM / Ejecta ~4±2 between the forward and reverse shocks with those swept-up from the surrounding interstellar medium and previous ejecta in the early Sedov-Taylor expansion, it appears that their sputtering and scattering interactions inject their refractories, and implanted volatiles, as suprathermal ions into those cosmic-ray accelerating shocks.

As much as ~0.45 $M_{\odot}$ of such cold, ~20 to 30 K, fast refractory grains, predominantly carbon and silicates, have been observed (Matsuura et al. 2011, 2015; Dwek & Arendt 2016) at velocities of ~2,000 to 3,000 km s$^{-1}$ (Kozasa et al. 1991; Colgan et al. 1994) in the freely expanding ejecta of the supernova, SN1987A, as it cooled within just two years after its explosion. Comparable surviving dust masses have also been found in much older core collapse remnants, such as the ~440 year old Cas A of 1680 (De Looze et al. 2017) and SNR G54.1+0.3 with an estimated age of 1500-3000 yr (Temim et al. 2017). Such large yields could in fact make core collapse supernovae the major source of interstellar dust (Dwek et al. 2007).

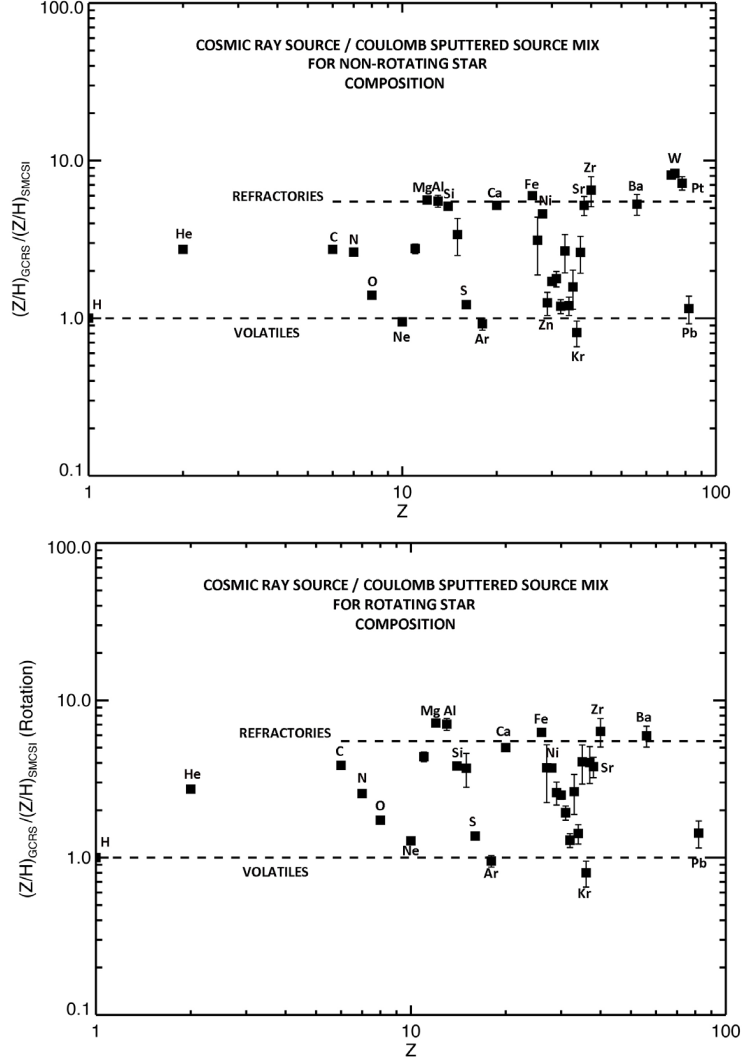

**Figure 6a,b**. The ratio of the cosmic ray source composition $(Z/H)_{GCRS}/(Z/H)_{SMCSI}$, relative to that of the source mix Coulomb sputtering injection, $(Z/H)_{SMCSI} = (Z/H)_{SM} F_{CS}$, where the Coulomb sputtering and scattering injection, $F_{CS} \sim Z/(1+Z^{2/3})^{1/2}$. This clearly reveals the last two remaining grain factors that determine the galactic cosmic ray source composition: a) the maximum refractory grain enrichment constant relative to volatiles, $C_{R/V} \sim 5.5$ (dashed line), and b) the various intermediate abundances determined by the fraction of each element that condenses in, or is implanted in grains, $F_{GC}$.

To clearly distinguish the contributions of the refractory/volatile grain constant $C_{R/V}$, and the grain condensation fractions, $F_{GC}$, from that of the Coulomb sputtering and scattering injection factor, $F_{CS}$, in Figure 6, we multiply the latter times the source mix elemental abundance distribution, $(Z/H)_{SM}$, to define the source-mix-Coulomb-sputtered composition, $(Z/H)_{SMCSI} = (Z/H)_{SM} F_{CS}$, resulting from the interactions of fast refractory grains with volatiles, primarily H and He, in the swept-up surrounding ISM. There may also be some significant accelerated refractory grains in that medium that interact with fast H and He in the ejecta.

Here we see that the relative injection enrichment by the fully condensed, highly refractory elements, such as Mg, Al, Si, Ca, Fe, Ni, Sr, Zr, & Ba, over the highly volatile ones, such as Ne,

S, Ar, Ge, Se, Kr & Pb, is approximately $C_{R/V} \sim 5.5$. This maximum grain sputtering injection factor is in good agreement with the ratio of 6.4±0.3 between volatile Ar, and close-by refractory Ca measured by Ogliore et al. (2009), since after correcting for the small Coulomb sputtering contribution of $(^{20}Ca/^{18}Ar)^{2/3} = 1.07$, this ratio is in fact 5.9±0.3. These values are higher than the maximum refractory/volatile enrichment ratios of ~3 to ~5 (Rauch et al., 2009, Fig. 9) and ~4 to ~4.5 (Murphy et al. 2016), based on differing refractory and volatile best-fit slopes for A-dependent enrichment, rather than a constant Z-dependence for both.

## 3.2 Refractory / Volatile Constant

The cosmic ray refractory/volatile enrichment constant, $C_{R/V} = 5.5\ F_{GC} + 1.0\ F_V$, brackets the range of partially condensed refractory grains, where $F_{GC}$ is the partially condensed grain fraction. From the Coulomb sputtering dependence of the noble gases, the remaining volatile fraction, $F_V = (1 - F_{GC})$ may also be interpreted as essentially the mean saturated implantation fraction for noble gases, or $\sim 1/5.5 \sim 0.18$. That would also appear to be consistent with the base abundance fractions of such volatiles in C3 chondrules indicated in Figure 7, and with extensive calculations (Marhas & Sharda 2018) of relative implantation fractions versus condensation of < 0.25 both Fe and Ni for the most probable models of 15-25 $M_\odot$ SN II CCSN grain formation.

Thus, $C_{R/V} = 5.5\ F_{GC} + 1(1 - F_{GC}) = 1 + 4.5 F_{GC}$. The elemental grain condensation fractions, $F_{GC}$, have been determined both indirectly from optical determinations (e,g. Savage & Sembach 1996; Jenkins 2009) of the average depletion of elements in the interstellar gas observed along various lines of sight through the interstellar medium compared to their solar system abundances, and directly from analysis (e.g. Wasson & Kallemeyn 1988) of early solar system and presolar grains, or type C chondrules, in carbonaceous chondritic meteorites. For each element these fractions show a clear correlation with the condensation / sublimation temperature of its refractory compounds (Kallemeyn & Wasson 1981; Wasson & Kallemeyn 1988; Savage & Sembach 1996; Lodders 2003; Jenkins 2009), as can be seen in Figure 7. The refractory elements in the most common carbonaceous chondrules, type C1, are thought to be fully condensed, $F_{GC} = 1$ for condensation temperatures $T_C >$ few 100 K. These relative abundances are consistent with solar photospheric values, and together they are taken (e.g Anders & Grevesse 1989; Lodders 2003) as the two best measures of solar system abundances, as well as local ISM abundances.

Detailed calculations (Bianchi & Schneider 2007; Nozawa et al. 2007; Sarangi & Cherchneff 2015; Bocchio et al. 2016; Biscaro & Cherchneff 2016; Micelotta et al. 2016) of dust formation and survival in core collapse supernova ejecta model the condensation of ~0.1 to 0.6 $M_\odot$ in refractory grains at ~20 K, consistent with observations mentioned above. They also suggest a surviving mass of the order of >0.1 $M_\odot$ in refractory grains after the destructive passage of reverse shocks, expanding in ambient densities $n < 0.1$ H cm$^{-3}$, running roughly proportional to $\sim n^{-1/2}$. During that passage, grains of radius 10 to 200 A are calculated (Bianchi & Schneider 2007) to reach temperatures between about 100 and 1000 K.

Such a grain temperature range is consistent with recent observations of grains in supernova remnants (Matsuura et al. 2015; Micelotta et al. 2018; Sarangi et al. 2018; and others noted above), as well as the interstellar medium (Savage & Sembach 1996). These temperatures are also quite consistent with the range of grain exposure temperatures between ~400 K and ~1400 K (see Figure 7), inferred from the partial condensation or sublimation fractions of carbonaceous chondrules of types C1 through C3 (Kallemeyn & Wasson 1981; Wasson & Kallemeyn 1988). We might expect that the highest temperatures that determined the partial grain sublimation fractions for both the ejecta and swept-up dust would be that in the hot, turbulent, supernova-shocked regions where the ejecta and swept-up interstellar grains are mixed and accelerated, and could therefore be essentially the same, $F_{GC}$ for both.

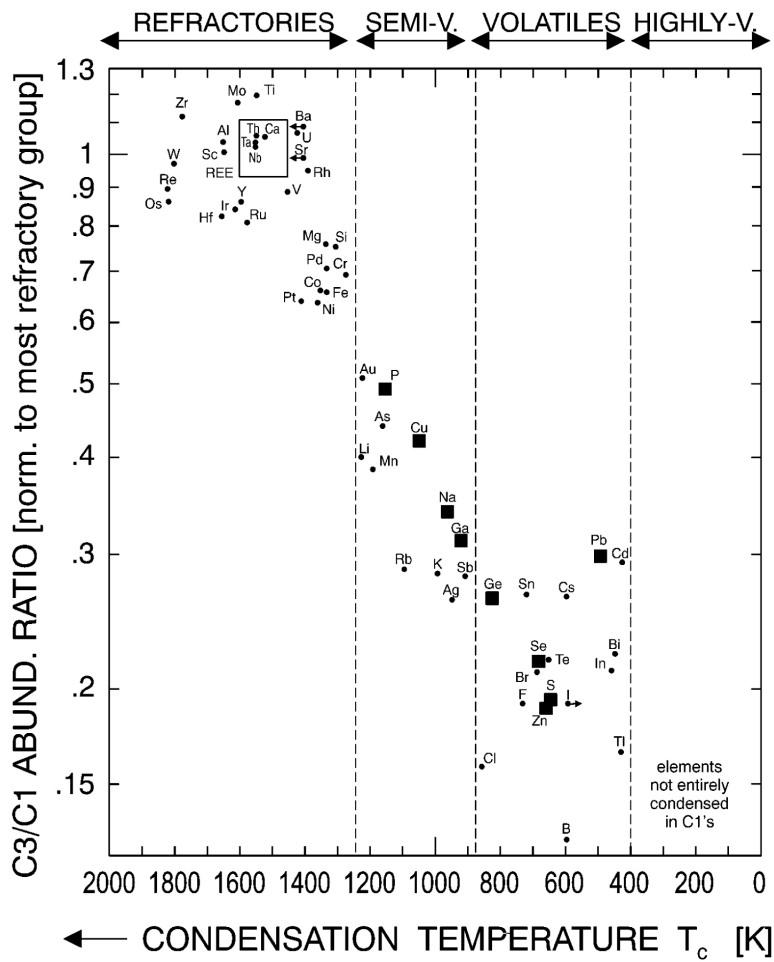

**Figure 7.** The partial condensation or sublimation fractions, $F_{GC}$, of each refractory element in carbonaceous chondrules, based on the ratio C3/C1 of that element in the hottest, least condensed or most sublimed grains, the C3 chondrules, to that in the coldest, most condensed, the C1 chondrules, measured by (Wasson & Kallemeyn 1988), shows an exponential correlation with the calculated condensation temperature (Kallemeyn & Wasson 1981; plotted by Meyer et al. 1997). Overall the fluctuations, $\Delta F_{GC} \sim 0.1$, but the relative variations $\Delta F_{GC}/F_{GC}$ approach 0.5 for the volatile and chemically active elements with condensation temperatures below 900 K.

Not only are core collapse supernovae thought (Dwek 1998) to be the main source of silicate grains in the interstellar medium, there is also clear evidence of a direct connection between the supernova grains and some of the presolar grain (Wasson 2017) inclusions in the C3 chondrules, particularly in calcium-aluminum inclusions (CAIs), which are the oldest known solids in the solar system. Isotopic anomalies in these inclusion, reveal the presence of short-lived radioactive nuclei with half-lives < 15 Myr: $^{26}$Al, $^{36}$Cl, $^{41}$Ca, $^{53}$Mn, $^{60}$Fe, $^{107}$Pd, $^{129}$I, $^{182}$Hf, and $^{244}$Pu, all of which are attributed to core collapse supernovae and their Wolf-Rayet winds (Adams 2010; Fujimoto et al. 2018, and the references therein). Although these radionuclei were first thought to have resulted from a rare, chance occurrence of a nearby supernova, they are now recognized as the natural result of the fact that the Sun, like nearly all stars, was formed in a highly spatially and temporally clustered star formation region. Moreover, the injection of these freshly synthesized nuclei into the proto-solar nebula is thought (e.g. Connelly et al. 2008; Goodson et al. 2016) to be via grains like the CAIs.

We suggest, therefore, that the temperature-dependent fraction of each refractory element that condensed and survived sublimation in the grains measured in the most heated sample, the class C3 of meteoritic chondrules, may also provide a reasonable first estimate of $F_{GC}$ for both the fast ejecta grains and the swept-up interstellar grains, since their different relative elemental abundances do not appear to have a major effect on the principal condensates. Nonetheless, the meteoritic fractions can also be significantly modified by subsequent heating, or cooling in the proto-solar nebula, particularly those of the chemically active and more volatile elements, such as Pb and Cd with condensation temperatures below ~ 900 K, which show (Figure 7) fluctuations of as much as a factor of ~2 above the relatively tight thermal correlation of those at higher temperatures and may result from subsequent recondensation or accretion (e.g. Snow 1975).

For O, in fact, the fraction in grains appears to be primarily determined (Meyer et al. 1997; Lingenfelter & Higdon 2007) by the sum of the fractions of O bound in highly refractory $SiO_2$, $MgO$, $Fe_3O_4$, $Al_2O_3$ and $CaO$ weighted by their relative source mix abundances: 1.48: 1.67: 1.21: 0.21: 0.07, which gives a grain $F_R \sim 0.15$ relative to the O abundance.

If the temperature and corresponding elemental grain condensation fractions $F_R$ for both fast ejecta grains and swept-up interstellar grains are indeed the same, then the grain condensation / sublimation constant, $C_{R/V} = 5.5\ F_{GC} + 1(1 - F_{GC}) = 1 + 4.5 F_{GC}$, times the grain Coulomb sputtering factor, $F_{CS} = Z / (1 + Z^{2/3})^{1/2}$, define the overall elemental grain injection factor, $F_{GI} = C_{R/V} F_{CS}$. This factor operates on the supernova source mix abundances to produce the final grain-injected source mix elemental abundances,

$$(Z/H)_{SMGI} = F_{GI}\ (Z/H)_{SM} = (1 + 4.5 F_{GC})\ [Z / (1 + Z^{2/3})^{1/2}]\ (Z/H)_{SM}, \qquad (2)$$

which can be directly compared with the observationally determined Galactic cosmic ray source elemental abundances, $(Z/H)_{GCRS}$.

### 3.3 Galactic Cosmic Ray Source Composition / Grain-Injected Supernova Mix

With the core collapse supernova mix of the ejecta and the interstellar medium, $(Z/H)_{SM}$, we can define the overall supernova source mix and grain enrichment, $(Z/H)_{SMGI} = F_{GI}(Z/H)_{SM}$, which is the expected supernova cosmic ray abundances. Comparing this model distribution with the cosmic ray source distribution determined from local cosmic ray abundance measurements $(Z/H)_{GCRS} / (Z/H)_{SMGI}$, we see from Figure 8, that the model is in good agreement, ~ ±35%, with the cosmic ray source abundances. This agreement also clearly suggests that the refractory grain condensation / sublimation fraction $F_R$ from the C3/C1 chondrules (Figure 7) gives, in fact, a reasonable estimate of both the ejecta and swept-up grain fractions.

Thus, we see that the extensive measurements of the local cosmic ray elemental and isotopic abundances from H to U have now shown very clearly that they differ radically from those of the solar system and local interstellar medium by factors varying from ~ 20 to ~200. But the nature of these differences have also revealed their primary causes: the mixing of supernova ejecta with the interstellar medium, and the injection of these nuclei as suprathermal ions sputtered from refractory grains into the supernova shocks, where diffusive shock acceleration carries them to cosmic ray energies.

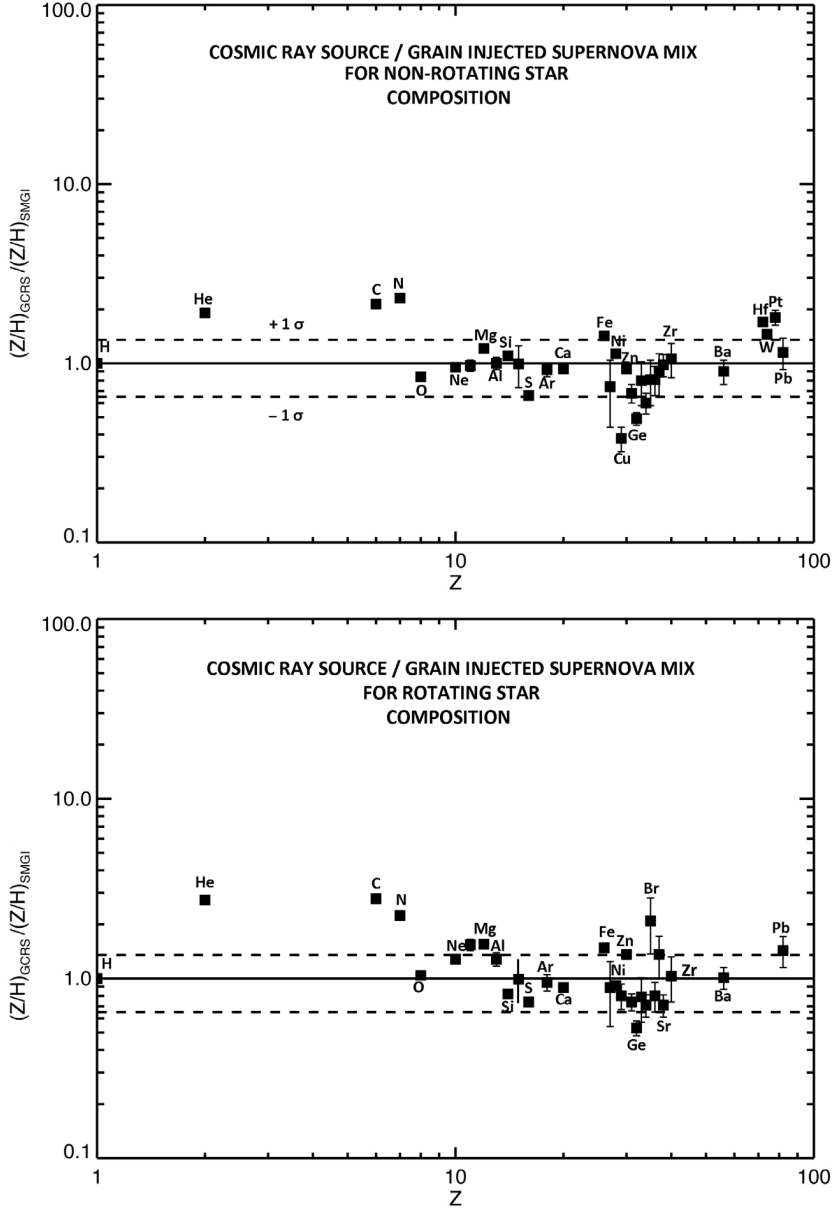

**Figure 8a,b.** The ratio of the observationally determined cosmic ray source elemental composition $(Z/H)_{GCRS} / (Z/H)_{SMGI}$, relative to the Grain-Injected Source Mix of the dominant s-process core collapse ejecta elemental composition, $(Z/H)_{SMGI} = F_{GI}(Z/H)_{SM}$. Applying the grain injection factor, $F_{GI}$, to the source mix, the factor of 5.5 spread in the variable grain fraction, shown in Figure 6, collapses to give remarkably good agreement in the elemental abundances between the grain-injected cosmic ray source mix $(Z/H)_{SMGI}$ and the galactic cosmic ray source values $(Z/H)_{GCRS}$ at a ratio of 1.0±0.35 at the ±1σ level. The conspicuous outliers that lie well above unity suggest underproduction of cosmic rays by the core collapse mix model and could indicate contributions of Wolf-Rayet winds, asymptotic giant branch (AGB) stars, SN Ia supernovae, and r-process nucleosynthesis that are discussed in the next sections.

Moreover, these new measurements have at last quantified these two basic processes in the generation of the cosmic ray composition. In particular, they show that the cosmic ray source elemental abundance ratios can be simply defined by a single mixing ratio by *mass* of high metallicity supernova ejecta and swept-up interstellar medium, coupled with the calculated s-process nucleosynthesis yields of the dominant core collapse supernovae. This mixing increases the elemental cosmic ray source abundances by factors of ~ 2 to ~10 over those of the interstellar medium.

The accompanying grain injection, which provides even larger enrichment, can be defined by the fractions of individual refractory elements that condensed into fast grains in the freely expanding ejecta, as estimated from measurements of meteoritic chondrules, which can account for additional factors of as much as ~ 5.5, and combined with their sputtering and scattering that injects them as suprathermal ions into the cosmic-ray accelerating shocks explains their strong elemental charge dependent enrichment of ~4 to ~20, proportional to $Z^{2/3}$, which reflects their relative Coulomb sputtering rates. We have shown that applying these two basic processes of mixing and injection to interstellar values can thus give source abundances of the major cosmic ray elements to within ±35% at a ±1σ uncertainty. This vastly reduces the net differences of cosmic ray source abundances compared to solar system abundances by factors of ~20 to ~200, and all with no free parameters.

3.4 CCSN Ejecta & ISM Mixing, Grain Condensation & Sputtering Injection Equations

In summary, we have shown that the elemental (Z) abundances of the cosmic ray metals at their source *(Z/H)$_{GCRS}$*, determined from local flux measurements after correcting for the effects of differential Z-dependent propagation and spallation, can be well modeled by two factors, a supernova source mix, *(Z/H)$_{SM/SS}$*, and a grain sputtering injection, $F_{GI}$, applied to the solar system abundances *(Z/H)$_{SS}$*, namely

$$(Z/H)_{SMGI} = F_{GI}(Z/H)_{SM/SS}(Z/H)_{SS}, \qquad (3)$$

where the source mixing factor,

$$(Z/H)_{SM} = (F_{EJ}(Z/H)_{EJ/SS} + [1 - F_{EJ}](Z/H)_{ISM/SS})(Z/H)_{SS}. \qquad (4)$$

This enriches the individual elemental cosmic ray abundances by factors of ~2 to ~10 times that of the solar system values, and depends both on the fluid dynamical process of the bulk mass mixing of the ejecta and swept-up interstellar material together with the underlying individual nuclear reaction processes generating the relative abundances of each element.

$F_{EJ} \sim 0.2$, is the best-fit, single mixing *mass* fraction of highly enriched supernova ejecta mixed with a swept-up interstellar mass fraction, and $F_{ISM} = (1 - F_{EJ})$, determined both from Be and B measurements of old halo stars (Duncan et al. 1992; Ramaty et al. 1997; Alibes et al. 2002; Tatischeff & Gabici 2018) and cosmic ray source abundances (Higdon & Lingenfelter 2003; Binns et al. 2005; Lingenfelter et al., 1998; Rauch et al. 2009; Ahn et al. 2010; Murphy et al. 2016). *(Z/H)$_{EJ/SS}$* are the individual elemental abundances of the combined pre-supernova winds and supernova ejecta relative to solar system values, integrated over a Salpeter initial mass function, calculated from s-process yields (Woosley & Heger 2007; Limongi & Chieffi 2018) for core collapse supernovae that make up the bulk (81%, Li et al. 2011) of Galactic supernovae, *(Z/H)$_{ISM/SS}$* = 1.32 (Z/H)$_{SS}$, are determined from Galactic chemical evolution models (Timmes et al. 1995); and *(Z/H)$_{SS}$* are the elemental abundances in the solar system (Lodders, 2003), primarily from fully condensed C1 chondrtic meteorites. The highly effective suprathermal ion injection factor,

$$F_{GI} = (1 + [(C_{R/V}) - 1]F_{GC})(Z/[1 + Z^{2/3}]^{1/2}), \qquad (5)$$

further enriches the individual elemental cosmic ray abundances by even greater factors of ~4 to ~100, and depends on both the individual elemental refractory grain condensation and sublimation temperature, and the elemental charge, $Z$, dependent grain sputtering and scattering yields. $C_{R/V}$ is a constant equal to the enrichment of the most refractory elements compared to the most volatile, equal to ~5.5 as shown above, while $F_{GC}$, are the grain temperature dependent fractions of each element that condensed from the ejecta gas into grains, determined from the elemental abundance ratios C3/C1 of the most heated and sublimed versus the most cooled and condensed chondrule measurements (Wasson & Kallemeyn 1988). Finally, $Z / [1 + Z^{2/3}]^{1/2}$ is the classic Coulomb sputtering cross section calculated from interactions between H and refractory grain nuclei of charge Z (Sigmund 1969, 1981).

The ability of these processes to explain within ±35% the bulk of the major cosmic ray metal abundances in terms of s-process nucleosynthesis in the dominant core collapse supernovae also provides a powerful new tool for determining the contributions of the Wolf-Rayet winds, the rarer thermonuclear SN Ia supernovae, the asymptotic giant branch stars, and the relative production of r-process elements by core collapse supernovae and neutron star mergers.

Last, we discuss the broader implications of these results on the cosmic ray source and sites, and their injection and acceleration, and show that they can provide a new self-consistent framework that ties together the Galactic sources, injection, acceleration, and propagation of the cosmic rays.

## 4. COSMIC RAYS FROM WOLF-RAYET WINDS, SN Ia & r-PROCESS IN CCSNe

The comparison of the observationally determined cosmic ray source compositions $(Z/H)_{GCRS}$ with that of cosmic rays generated by the s-process elements in the core collapse supernovae ejecta, $(Z/H)_{SMGI}$, shows that this Grain-Injected-Ejecta-Mix model can account for most of the cosmic ray elemental compositions to within ±35% at ±1σ. In Figures 8a and 8b, however, there are several elemental abundances, i.e. Na, Mg, Fe, Cu, Ge, Se, Br, Hf, & W, among the ~ 1/3 of the values that could be expected to randomly lie above, or below, ±1σ.

But there are also a few very conspicuous outliers in this comparison that appear to signal contributions from other sources. As noted above, those outliers that lie well above unity suggest significant underproduction of cosmic rays by the core collapse s-process model, and indicate possible contributions from extra sources of $(Z/H)_{XMI} = [\{(Z/H)_{GCRS}/(Z/H)_{SMGI}\} - 1] (Z/H)_{SMGI}$. In particular, the cosmic ray C&N over abundances compared to expected CCSN yields suggests an extra contribution from Wolf-Rayet winds or asymptotic giant branch (AGB) stars. Other scattered outliers above (Z > 30), particularly Pt, may result from extra contributions of r-process elements from CCSNe not calculated along with the s-process by (Woosley & Heger 2007). Perhaps most important, there is a surprising lack of an Fe outlier expected from SN Ia supernovae, which shows a clear lack of acceleration of cosmic ray metals by these supernovae.

### 4.1 C & N

We consider first the > 3 σ outliers and strong indicators of underproduction of cosmic ray C and N by the core collapse supernova grain-injection source-mix models, both nonrotating and rotating (Figure 8). The C and N ratios $(Z/H)_{GCRS}/(Z/H)_{SMGI}$ do not equal a value close to 1.0, as would be expected solely from these models. Instead, for the nonrotating model $(Z/H)_{GCRS} / (Z/H)_{SMGI}$ ~ 2.14 ± 0.06 ± 0.35 for C, and ~ 2.25 ± 0.12 ± 0.35 for N, and for the rotating model ~ 2.78 ± 0.08 ± 0.35 for C, and ~ 2.24 ± 0.12 ± 0.35 for N or an overall mean of ~ 2.2 (see Appendices A & B). This underproduction suggests that there is an extra source of cosmic rays $(Z/H)_{XMI}$, such that $(Z/H)_{GCRS} = 2.2 (Z/H)_{SMGI} = 1.0 [(Z/H)_{SMGI} + (Z/H)_{XMI}]$, or simply that *(Z/H)$_{XMI}$ ~ (2.2 – 1.0) (Z/H)$_{SMGI}$ ~ 1.2 ± 0.35 (Z/H)$_{SMGI}$ at 3.4 σ*, that contributes about 1.2/2.2 ~ 55 ±16% of both the cosmic ray C and N.

Both AGB stars and Wolf-Rayet winds appear to be capable of producing such an amount of C and N. It has long been argued (e.g. Timmes et al. 1995; Dwek 1998; Gavilan et al. 2005) that the lower and intermediate mass, 2 < M$_\odot$ < 5, asymptotic giant branch stellar winds could be the major source of C and N in the interstellar medium, producing as much as 80% of the C (Mattsson 2010). Similarly, Wolf-Rayet winds of the most massive stars, ~ 40 < M$_\odot$ < 120, dredging up incompletely burned s-process elements, have also been suggested (Meyer et al. 1997, Fig. 6) to contribute between 50% and 80% of the cosmic ray C, accelerated by supernova shocks. Wolf-Rayet winds would also appear to be the source of the factor of 5.0±0.2 enrichment of $^{22}$Ne in the cosmic rays (Binns et al. 2001; Higdon & Lingenfelter 2003; Binns et al. 2005), although the yield is still uncertain.

The elemental production abundances of the extra source, *(Z/H)$_X$*, alone, however, do not determine what fraction of that *(Z/H)$_{XMI}$* is accelerated to cosmic ray energies. That fraction depends on whether or not the elements from the extra source are injected into the supernova shocks by the strongly biased grain condensation and sputtering process, such as *F$_{XI}$ ~ F$_{GI}$*, or by a relatively unbiased process, e.g. *F$_{XI}$ ~ 1*. Also it depends on what fraction, *F$_{XC}$*, of the extra source material can both spatially and temporally interact with supernova shocks. Although we also expect that the fresh wind material will mix with its surrounding material, we do not expect any significant direct wind shock acceleration comparable with that of supernovae, and assume that any fresh wind material is simply included the interstellar material swept-up by supernova shocks.

Since grains condense in the freely expanding and cooling winds of Wolf-Rayet, AGB, and other stars (e.g. Dwek 1998), we assume that the grain partial condensation / sublimation fractions of C and N from the winds of both Wolf-Rayet and AGB stars that might interact in supernova shocks are roughly the same as those fractions in the core collapse mix, as are their Coulomb sputtering factors, and thus that for both extra sources, *F$_{XI}$ ~ F$_{GI}$*.

The C and N in the Wolf-Rayet winds are also closely linked with dominant SN II core collapse supernova even though they are not their progenitors. The Wolf-Rayet phases of the most massive stars, ~ 40 to ~120 M$_\odot$, are calculated (e.g. Schaller et al. 1992) to last as much as ~ 6 Myr, which is just a few Myr before the most massive ~ 20 to 25 M$_\odot$ SN II core collapse progenitors begin to explode at an age of ~ 7 to 9 Myr, so their supernova shocks might easily interact with local C and N from Wolf-Rayet winds in a tight star formation cluster with an interacting fraction *F$_{WRC}$ ~ 1*.

On the other hand, since even the largest ~ 5 M$_\odot$, fastest evolving, AGB stars do not reach their giant phase until they are ~ 100 Myr old, while the smallest ~ 8 M$_\odot$, longest lived, core collapse supernova progenitors explode at an age of ~ 30 Myr (e.g. Schaller et al. 1992), it seems unlikely that the blast waves of the latter would encounter any concentrated ABG wind material even in a tight, roughly coeval, star formation cluster. Therefore, we would expect that their interacting fraction *F$_{AGBC}$ << 1*. Therefore, the Wolf-Rayet winds would appear to be by far the most likely source of cosmic ray C and N, if they can be shown to provide the extra ~ 1.2 times that of the core collapse supernova, which is needed to explain the cosmic ray abundances.

### 4.2 Missing SN Ia Cosmic Rays

In stark contrast to this extra source contribution is the lack of any comparable component of the cosmic ray Fe, or other metals, that were expected from the SN Ia supernovae. Calculations of Galactic Fe chemical evolution, driven primarily by ejecta of both CCSN of young, massive stars and thermonuclear SN Ia of older, accreting white dwarfs (e.g. Timmes et al. 1995; Prantzos et al. 2018), suggest that the SN Ia have produced anywhere from a third to two-thirds of the total $^{56}$Fe in our Galaxy, or roughly 1.25±0.75 times as much $^{56}$Fe as that of the CCSN.

But that total Galactic $^{56}$Fe production over the age of the Galaxy is strongly biased toward the higher early CCSN, while the recent calculations of the current $^{56}$Fe nucleosynthetic yields, reviewed by Chieffi and Limongi (2017), suggest that CCSN yields range from about 0.07 to 0.13 $^{56}$Fe M$_\odot$/SN, compared to a SN Ia yield of about 0.6 to 0.7 $^{56}$Fe M$_\odot$/SN (Iwamoto et al. 1999). Similarly, the most recent analysis of the light curve of the CCSN 1987A gives a parent $^{56}$Ni mass of 0.071±0.003 M$_\odot$ (Seitenzahl et al. 2014), while analysis of the light curve of the Ia SN 2013aa gives a corresponding $^{56}$Ni mass of 0.732±0.151 M$_\odot$ (Jacobson-Galan et al. 2018) quite consistent with the expected nucleosynthetic yields. Taken together with the relative rates of CCSN 2.30±0.48 SN/100 yr to SN Ia 0.54±0.12 SN/100 yr expected in our Galaxy, based on extragalactic observations (Li et al. 2011), the present Galactic $^{56}$Fe production rate by SN Ia is ~ 0.395±0.087 and by CCSN ~ 0.163±0.034 M$_\odot$/100 yr with SN Ia producing ~2.42±0.74 times as much Galactic $^{56}$Fe as CCSN.

From these estimated yields it might indeed be expected that the cosmic ray *(Fe/H)$_{GCRS/SMGI}$* could be ~ *1 + (Fe/H)$_{SNIaMI}$ / (Fe/H)$_{SMGI}$* ~ *1 + 2.42±0.74* ~ *3.42±0.74*, an even greater outlier than those of C or N, *if* the injection factors and the source mixing ratios were the same for SNIa and CCSN. But that is clearly not observed. Instead of the expected *(Fe/H)$_{GCRS/SMGI}$* being *3.42±0.74 (Fe/H)$_{SMGI}$*, an unmistakable ~ *7±2* σ above the CCSN grain-injection source mix model value of 1.0±0.35, we see from Figure 8a,b (and Appendices A & B) that *(Fe/H)$_{GCRS}$* ~ *1.45±0.35(Fe/H)$_{SMGI}$*, a potential excess of 0.45±0.35, only ~ *1.3* σ, suggests *no* significant overproduction of cosmic ray Fe from SNIa, or any other sources.

But this is not really so surprising in view of the now demonstrated (Figure 5) dominance of grain injection on cosmic ray refractory abundances, coupled with the observational failure to find "any significant mass of dust" in the ejecta of known SN Ia remnants (Dwek 2016, and the reference therein). For even though massive amounts of cold, freshly condensed dust grains have been found in young CCSN remnants, as discussed above, extensive searches for such newly-formed grains have found nothing, despite the massive amounts of iron produced in the relatively young SN Ia remnants, Tycho (1572), SN1006 RCW 86 (185 AD), and N103B in the LMC.

Without the injection of ejecta metals (Z > 2) as suprathermal ions into the accelerating shocks, the expected contribution of SN Ia to the cosmic ray metals is reduced relative to CCSN by the very large grain injection factor, $F_{GI} = (1 + [(C_{R/V}) – 1]F_{GC}) (Z / [1 + Z^{2/3}]^{1/2})$. For Fe, lacking the refractory / volatile enrichment of ~ 4x and the ~$Z^{2/3}$ Coulomb sputtering injection enrichment of ~8, yields a total factor of ~32 reduction in the expected extra cosmic ray Fe contribution of SN Ia ejecta, cutting it from ~2.4 times that of CCSN to a truly negligible 0.08, compared to CCSN. Thus the only SN Ia contribution to Fe in the cosmic rays appears to be that accelerated in CCSN remnant from their swept-up ISM component.

There is of course no similar suppression of cosmic ray volatile, H & He acceleration by SN Ia, as has been shown from the high energy, 100 MeV to 30 TeV, pion-decay gamma ray emission from the Tycho remnant (e.g. Morlino & Caprioli 2012; Lingenfelter 2018).

Thus it appears that the lack of a significant SN Ia contribution to cosmic ray Fe and other metals clearly results from their notable lack of ejecta grains (e.g. Dwek 2016) and their subsequent lack of strong grain injection and enrichment relative to that of the dusty core collapse supernovae. This linkage of the absence of ejecta grains and the absence of a cosmic ray metal component provides further evidence of the importance of grain injection in core collapse supernovae and makes them essentially the only accelerators of comic ray metals.

### 4.3 CCSN r-Process

In addition to the s-process nucleosynthesis, the elements of Z >30 can also be made by the explosive r-process in both CCSN and the newly observed neutron stars mergers (NSM). But the relative yields of these two sources are unknown. For the cosmic ray abundances, however, the CCSN fraction may at last be measurable. Since it appears that the cosmic ray metals almost all come solely from the CCSN source mix, only the CCSN s-process and r-process fraction plus that in the measured contribution from the swept-up ISM, will be accelerated to cosmic ray energies.

Moreover, comparisons of cosmic ray elemental abundance ratios measurements (Combet et al. 2005; Donnelly et al. 2012) of the ThU/Pt group with supernova r-process nucleosynthetic calculations (Lingenfelter et al. 2003) also suggest a similar ~20% ejecta mixing ratio $F_{EJ}$ for such products of explosive r-process supernovae. The possibility of determining the relative contributions of CCSN s- and r-process yields can also be seen from the $(Z/H)_{GCRS/SMGI}$ outliers above (Z > 30) that appear to result from the contributions of r-process elements to the cosmic rays from CCSNe in addition to that calculated by (Woosley & Heger 2007) solely for the s-process. For cosmic ray Pt the $(Z/H)_{GCRS} \sim 1.80\pm0.35\ (Z/H)_{SMGI}$ (Figure 8) suggests a CCSN s-process underproduction and a residual CCSN r-process contribution of ~ $0.80\pm0.35$ times that of the s-process. Thus we are presently analyzing all of the well-measured cosmic ray r-process elements to deduce similar r-process contributions for each, and reduce the net error of the mean CCSN fraction.

## 5. IMPLICTIONS OF SOURCE MIXING & GRAIN INJECTION ON COSMIC RAY SOURCES & SITES

Here we focus on the newly demonstrated link between the ubiquitously measured cosmic ray source mass mixing ratio of supernova ejecta to swept-up interstellar medium, and the homologously evolving supernova remnant shocked mix at the onset of adiabatic, Sedov-Taylor expansion. This uniquely defines at last the source, time and place of peak cosmic ray acceleration within that evolution. We also consider the dominance of grain sputtering and ion injection on the cosmic ray elemental composition variations, and the overall efficiency of diffusive shock acceleration in supernova remnants. This places critical constraints on the source environments and sites of such cosmic ray acceleration because of the major destruction of such grains by reverse shocks at that same time in the Sedov-Taylor expansion.

But first we briefly review the nature of the sources and distribution of supernovae in our Galaxy and the basic nature of their evolution. Most supernovae do not occur randomly throughout the interstellar medium. Instead, their progenitors, which account for just ~0.1%, of all stars, are born in episodic starbursts in the densest parts of giant molecular clouds that contain up to $\sim 3 \times 10^6$ $M_\odot$ of gas and dust (e.g. Pudritz 2002: Branch & Wheeler 2017). There many thousands of stars are formed within a Myr in tight clusters of only a few pc in diameter. The most massive and most rapidly evolving of these stars, the 15-120 $M_\odot$ O stars (Martins et al. 2005) and the 8-15 $M_\odot$ brightest B stars, the red supergiants (RSG), end their lives as core collapse supernovae. Moreover, they pass on much of their natal stellar clustering from their compact OB associations. With a mean radial dispersion velocity of only ~2 km s$^{-1}$, or ~2 pc

Myr$^{-1}$ (de Zeeuw et al. 1999), most of these supernova progenitors disperse <50 pc (Higdon et al. 1998; Higdon & Lingenfelter 2005) during their relatively short lives of just 3 to 35 Myr before they become supernovae (Schaller et al. 1992; Chieffi & Limongi 2013). Thus, the core-collapse supernovae are also *highly correlated in both space and time*.

Such core collapse supernovae account (Li et al. 2011; Shivver et al. 2017) for ~81% of the Galactic total at a rate of 1 CCSN every ~43 yrs. That leaves the thermonuclear SN Ia from the far more slowly evolving, ~ Gyr or more (Greggio 2005), and hence quite widely scattered, accreting white dwarf binaries to produce the bulk of single supernova remnants in the interstellar medium at 1 SN Ia every ~180 yrs. There are of the order of 100 supernova progenitors in a typical OB association (McKee & Williams 1997; Williams & McKee 1997; Higdon & Lingenfelter 2005). Observations (e.g. Blaauw 1991) show that these are generally produced as part of a series of several bursts of star formation separated in space and time by ~50 pc and ~4 Myr, as their parent molecular cloud complexes form multiple OB associations.

These OB clusters are surrounded by self-generated, giant H II superbubbles of basically the local ISM, a prominent feature of Galactic star formation regions. Blown by the powerful UV radiation and the Wolf Rayet winds of the most massive O stars and the culminating supernovae, they formed cavities in the molecular clouds, blowing out most of the gas and magnetic field into surrounding warm ~$10^4$ K, dense ~10 H cm$^{-3}$, HI supershells, or cocoons. These enclose their merged, supernova-remnant, superbubble cavities of >$10^5$ pc$^3$ filled with hot >$10^6$ K, tenuous ~0.001 to <0.01 H cm$^{-3}$ gas (e.g. Cox & Smith 1974; McKee & Ostriker 1977; Weaver et al. 1977; McCray & Snow 1979; Heiles 1987, Mac Low & McCray 1988; Shull & Saken 1995; Kim & Ostriker 2017).

Because of the asymmetry of the swept-up, compressed magnetic field pressure, the superbubbles tend to be roughly tubular in shape extending along the magnetic field (Tomisaka 1992) until their radii approach the scale height of the Galactic disk. Then they vent into the halo (Kim & Ostriker 2018), forming what Heiles (1979; 1994) dubbed "worms" and "chimneys" of hot superbubble gas, enclosed by supershells of warm ionized gas.

As few as 5 supernovae in a cluster appear to be all that is needed to generate a superbubble (Higdon & Lingenfelter 2005). So from the size distribution of the young star clusters, ~85% of the core collapse stellar progenitors are expected to be born in superbubble generating clusters. In that crowded environment, as many as half of those smaller OB clusters may also be engulfed by neighboring superbubbles.

In addition, it appears that essentially all of the core collapse supernova progenitors that missed being engulfed by superbubbles, still explode very close to some part of the parent molecular cloud complex, and can thus account (Lingenfelter 2018) for essentially all of the single gamma ray supernova remnants observed by *Fermi* to be interacting with filaments of molecular clouds. For, assuming a detectable lifetime (Acero et al. 2016) of anywhere from ~10 to 50 kyr, those 30 to 44 gamma ray remnants interacting with molecular gas, give an occurrence rate of ~1 to 4 SNR/kyr out of ~23 SNR/kyr for all Galactic core collapse supernovae. Independent of the *Fermi* sky coverage, this amounts to ~4 to 17% of core collapse supernovae, and from a quarter to all of the estimated ~15% that were born in some of the smaller clusters that were unable to create superbubbles of their own.

Core collapse supernovae in hot ~$10^6$ K superbubbles likely account (Higdon & Lingenfelter 2005) for ~75% of all Galactic supernovae, occurring at an overall rate of 1 CCSN every ~43 years with an average of ~300 supernova occurring in each of an estimated ~3,000 merged superbubbles over their ~50 Myr lifetimes. The supernovae in these low density $n$ <0.01 H cm$^{-3}$ superbubbles, do not produce many, small ~20 pc, scattered, individual supernova remnants like those in the cooler, denser >0.1 H cm$^{-3}$ phases of the interstellar medium that radiate away most of their shock energy in <50 kyr. Instead they can grow to ~100 pc. Moreover, these supernovae occurring in the bursts of highly clustered stars in the larger, more tenuous ~0.001 H cm$^{-3}$, can interact with older, nearby supernova remnants, their blastwave shocks may extend as much as a few hundred pc as the superbubbles of nearby associations merge and collectively (e.g. Bykov & Fleishman 1992; Bykov & Toptygin 2001) continue accelerating cosmic rays for ~50 Myr or more until their last supernova explodes.

Since ~300 such spatially and temporally clustered CCSN occur in a supperbubble at rate of about ~1 CCSN every ~150 kyr in its typical lifetime of ~50 Myr, not only do they each serially accelerate fresh ejecta and swept-up ISM by diffusive shock acceleration, these shocks also further accelerate some of the cosmic rays from previous supernovae.

So the extensive superbubble acceleration processes also appear (Bykov & Toptygin 2001; Ferrand & Marcowith 2010) to be able to account for both the acceleration and the escape spectra of the locally measured cosmic rays with the simple -2.7 power-law energy spectra with constant nuclear abundance ratios up to the so called "knee" above $10^6$ GeV. There the proton spectra breaks (Horandel 2013) at $E_p$ ~$4\times10^6$ GeV, followed successively at higher energies by the less abundant helium and heavier nuclei breaking at $E_Z \sim E_p Z$, where Z is their nuclear charge, all the way up to ~$4\times10^8$ GeV (Z~92) as the mean nuclear mass of the remaining cosmic rays increases. At the same time the further acceleration by the multiple supernova shocks help smooth the overall spectral index to about -3.1 above the knee on up to the "ankle" at ~ $10^9$ GeV.

The thermal X-ray emission from the hot tenuous cores of these superbubbles is generally too faint and diffuse to be resolved in all but some of the closest and largest bubbles (e.g. Cash et al. 1980). The surrounding, much denser HI supershells nonetheless are readily seen in 21-cm emission, and these and their enclosed voids led to their discovery (Heiles 1979). The youngest superbubbles are also revealed by much shorter lived, ~7 Myr, bright Hα emitting, HII shells on the inner walls of the cocoons, ionized by the intense far ultraviolet emission of the Wolf Rayet stars before they blow off most of their mass, collapse and explode as SN Ic (van Dyk et al. 1996; Anderson et al. 2012).

Most recently observations of high-energy gamma rays from decay of $\pi^o$, which are produced by relativistic *pp*-interactions and are a signature of cosmic ray proton interactions, have also shown that cosmic rays are being accelerated with a spectrum of ~ $E^{-2.3}$ up to the "knee" at a PeV in supernova remnants of spatially and temporally clustered CCSNe in such massive superbubbles as the Cygnus Cocoon, the Galactic Central Molecular Zone (CMZ), and both Westerlund 1 and 2 (Abdo et al. 2009; Ackermann et al. 2011; Aharonian et al. 2018).

The Cygnus Superbubble (Figure 9), the closest and best studied of these, is fed by at least ten OB associations and smaller clusters (Wright et al. 2015), and it has been growing steadily for at least 20 Myr (Comeron et al. 2016, and the references therein). In Cygnus OB2, the youngest and biggest of these associations, some 55 O stars of 15 M$_\odot$ or more have currently been identified (Wright et al. 2015), plus at least 88 early B0-B2 star RSG supernova progenitors of 15 to 8 M$_\odot$. Most of the O stars in Cyg OB2 were produced between 1 and 7 Myr ago, possibly peaked around 4 to 5 Myr ago (Wright et al. 2015). Since the lifetimes of the most massive of these stars are only ~3 Myr (e.g. Schaller et al. 1992; Chieffi & Limongi 2013), a number of these stars have already exploded and others have lost significant mass in their winds, so their mass distribution has evolved considerably. In addition, there are all the other older OB associations within the superbubble with ages of as much as ~20 Myr (Comeron et al. 2016). Together these suggest (Lingenfelter 2018) a mean supernova rate of ~6-9 CCSN/Myr that will produce some ~300 to ~500 core collapse supernovae over the expected ~50 Myr lifetime of just the Cygnus Superbubble alone.

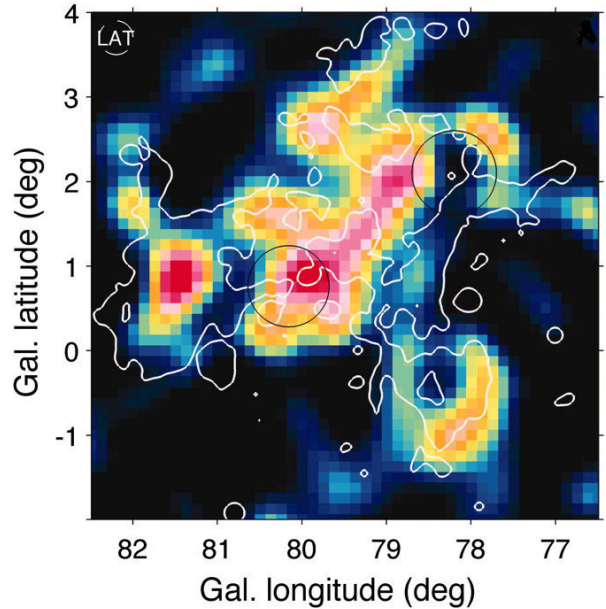

**Figure 9.** Cosmic ray produced $\pi^o$-decay $\gamma$-rays from the Cygnus Superbubble Cocoon detected by *Fermi* (Ackermann et al. 2011). Scale $4^o \sim 100$ pc.

High energy $\pi^o$-decay $\gamma$-rays have also shown (Morlino & Caprioli 2012) proton acceleration up to nearly the same energy in one of the three youngest SN Ia supernovae in our Galaxy (Tycho Brahe's of 1572), as well as from about 30 individual CCSN remnants, such as W44 and IC433, scattered throughout the interstellar medium (Albert et al. 2007; Ackermann et al. 2013). These latter supernova remnants, though not in superbubbles, all belonged to a small, complex class of "mixed-morphology" remnants (Rho & Petre 1998) that were interacting in dense $> 10$ H cm$^{-3}$ filamentary molecular clouds which made them much brighter than others in the average ~0.1--1 H cm$^{-3}$ interstellar medium. But, because of their much higher accompanying ionization losses, they were far less efficient, and their inferred cosmic ray proton spectra broke (Albert et al. 2007; Acciari et al, 2011) below $E^{-2/3}$ at barely 20 to 200 GeV, which is only ~0.01 to 1% of the PeV break energy of the cosmic rays in CCSN remnants in hot, low density $< 0.01$ H cm$^{-3}$ superbubbles (Lingenfelter 2018; Aharonian et al., 2018).

### 5.1 Supernova Remnant Evolution, Grains and Cosmic Rays

Tracing the evolution of the ejecta mixing and grain injection processes, we follow the general three-stage evolutionary approximations of supernova remnant expansion from free expansion to the Sedov-Taylor, or adiabatic stage, when after a time depending on the surrounding interstellar gas density, the ejecta sweeps up a comparable mass of interstellar matter, driving a strong blastwave, or forward shock into the ISM and in reaction a reverse shock into the ejecta that both can efficiently accelerate cosmic rays. The remnant continues in that stage until the radiative, or "snowplow" stage when so much more mass has been swept up that most of the remaining energy is radiated away and the expansion becomes subsonic.

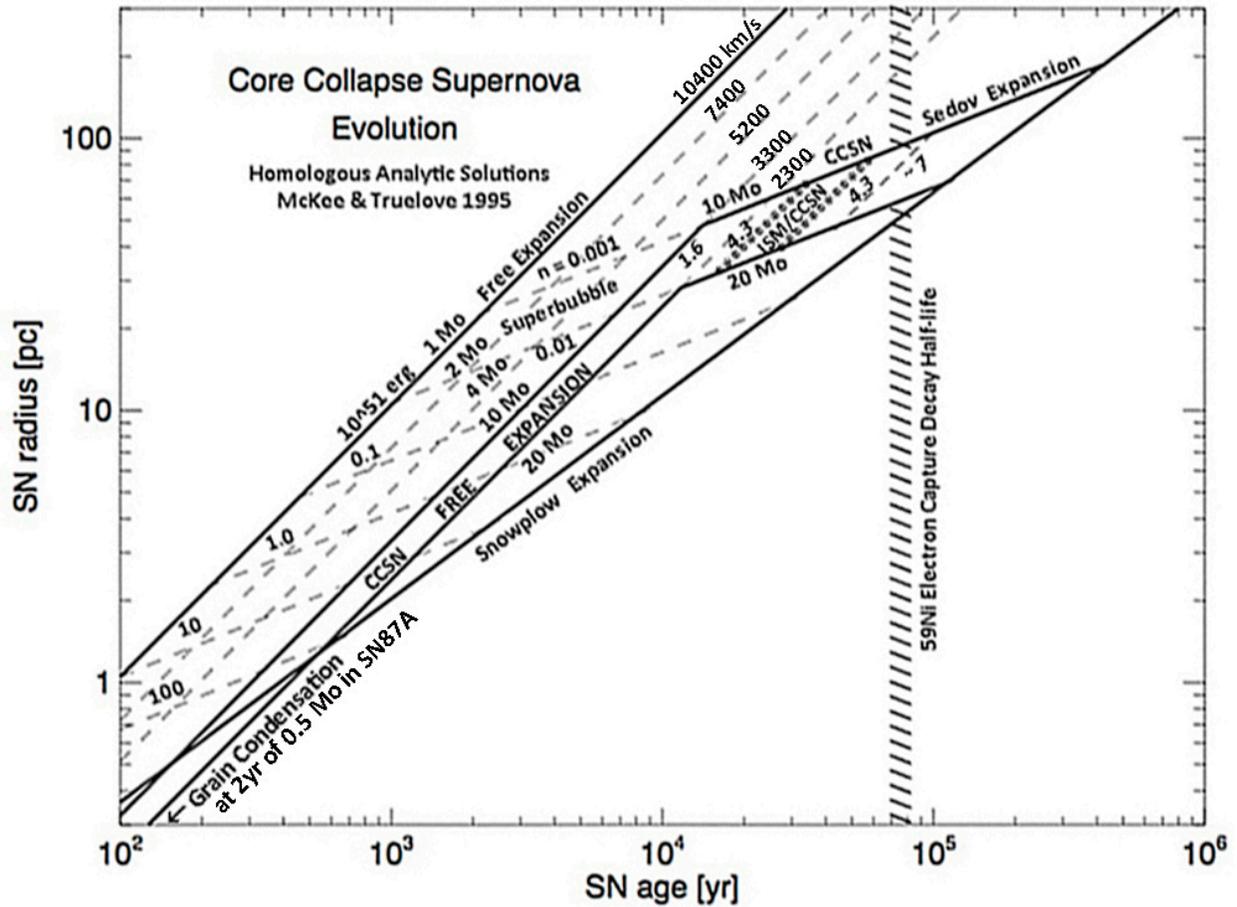

**Figure 10**. Homologous evolutionary tracks of Supernova Radius vs. Age as a function of ambient density, ejecta mass, and supernova energy (McKee & Truelove 1995): from ejecta-driven, Free Expansion with high-velocity Refractory Grain Formation, through Sedov-Taylor Expansion and CCSN Ejecta Mixing Ratio with swept-up ISM, turbulent Gas-Grain Sputtering, Suprathermal Ion Injection and Diffusive Shock Acceleration of cosmic rays with a measured (Murphy et al. 2016) mean mixing mass ratio of cosmic ray ISM/CCSN ~ 4.3, before the Forward shock eventually goes subsonic in radiative, or "Snowplow" Expansion.

In particular, we adopt McKee & Truelove's (1995) treatment of supernova remnant expansion, which demonstrates that all uniform nonradiative SNRs moving into uniform, homogeneous media can be described by a single, *universal* solution in terms of a characteristic remnant age, $t_{ST} = 209\ M_{EJ}^{5/6} E_{51}^{-1/2} n_o^{-1/3}$ yr, and radius, $R_{ST} = 2.23 M_{EJ}^{1/3} n_o^{-1/3}$ pc, at the effective onset of the Sedov-Taylor stage of expansion. These analytic expressions for the remnant radius versus age, and that for the velocity below, define the homologous evolution of the supernova remnant from the explosion to the end of the Sedov-Taylor expansion at the onset of pressure-driven snowplow (PDS) stage, $t_{PDS} \sim 13{,}300\ E_{51}^{3/14} n_o^{-4/7}$ yr and $R_{PDS} \sim 14.0\ E_{51}^{-2/7} n_o^{-3/7}$ pc. The supernova remnant evolutionary tracks drawn from these expressions are shown in Figure 10.

Immediately after a supernova explosion its high temperature ejecta remnant approaches a Free Expansion stage, quickly cools to <100 K in the moving frame, and the refractory elements condense into grains. Assuming that all supernovae produce the same blast energy of $E_{51} \sim 10^{51}$ erg, independent of their ejecta mass, $M_{EJ}$, the remnants are calculated (McKee & Truelove 1995) to be expanding freely at a velocity, $V_{FE} = R_{ST}/t_{ST} \sim 10400\, E_{51}^{1/2} M_{EJ}^{-1/2}$ km s$^{-1}$, or $\sim$ pc Myr$^{-1}$. For the CCSN remnants, whose ejecta masses lie primarily between $\sim$10 to 20 M$_\odot$, their free expansion velocities range between about 3300 to 2300 km s$^{-1}$. Thus, for the majority of supernovae, the CCSN whose progenitor O & B stars are both born and die in the uniform hot, tenuous superbubbles, where with densities ranging from $n \sim 0.01$ to 0.001 H cm$^{-3}$ (e.g. Mac Low & McCray 1987; Tomisaka 1992; Higdon et al. 1998; Higdon & Lingenfelter 2005), their remnants typically take about 10 to 15 kyr to expand to about $\sim$ 30 to 50 pc radius and enter the Sedov-Taylor stage of expansion. A fraction of the CCSN remnants (Lingenfelter 2018), such as those isolated *Fermi* high energy gamma ray SNRs that encounter denser >0.1 H cm$^{-3}$ interstellar gas, can even sweep up enough to enter the Sedov-Taylor expansion phase in <10 kyr. In all, the bulk mixing of the CCSN ejecta and swept-up ISM drives further Rayleigh-Taylor turbulent mixing, leading to gas-grain sputtering interactions, and suprathermal ion injection into the diffusive shock acceleration of a small fraction of the ejecta-ISM mix to cosmic ray energies.

### 5.2 Swept-up ISM to CCSN Ejecta Mass Mix

Here we discuss the telling implications of the ubiquitous $\sim$4 to 1, mixing ratio of swept-up ISM to CCSN ejecta mass, defined by analyses of the cosmic ray source composition in Section 2, and Figure 5.

We (Higdon et al. 1998) originally suggested that the cosmic ray mass mixing ratio might result from the build-up of an essentially constant ejecta-enriched interstellar material from the clustered CCSN in the cores of superbubbles. But as we now see in Figures 10 & 11, there is no constant ejecta fraction in the CCSN during the reverse shock passage. In every remnant, that fraction continuously decreases by $\sim t^{-6/5}$, diluted by more swept-up ISM, and reaching just a few percent as its radius approaches that of the galactic disk scale height of $\sim$100 pc at which superbubbles themselves vent their contents through "worms" and "chimneys" (e.g. Heiles 1987; Kim & Ostriker 2017).

Yet, as we also see in Figure 11 particularly, the OB-clustered CCSN explosions in the hot, low density superbubbles can, in fact, produce the uniform cosmic ray mass mixing ratio simply through the standard homologous evolution of supernova remnants. What's critical about superbubbles is their low density, <0.01 H cm$^{-3}$. That both allows the bulk of the grains to survive the passage of the reverse shock through the mixing ejecta during the early Sedov-Taylor phase of grand sputtering, injection and acceleration, and delays the major fraction of that acceleration until the expected cosmic ray $^{59}$Ni abundance is consistent with observed limits.

The latest analysis (Murphy et al. 2016) of the major cosmic ray source abundances from N through Zr gives a best-fit 1$\sigma$ value of the CCSN ejecta mass fraction in the cosmic ray source $F_{EJ}$ of $19_{-6}^{+11}$ %, or a mean swept-up ISM to CCSN ejecta mass mixing ratio ISM/CCSN = $(1 - F_{EJ})/F_{EJ}$ of $4.3_{-2.0}^{+2.4}$, spanning a broad $t^*/t_{ST}$ range from 2.3 to 6.7 at peak acceleration.

At the onset of the Sedov-Taylor stage, CCSN building strong forward and reverse shocks, had swept up and mixed with $\sim$ 1.6 times as much mass of ISM, as that of their own ejecta (McKee & Truelove 1995; Truelove & McKee 1999). This homologous evolving remnant mass mixing ratio of shocked ISM mass to SN ejecta would seem quite naturally to be roughly that of the relative mix of the source material that is accelerated to cosmic ray energies by those shocks.

The bulk mixing mass ratio of this growing swept-up ambient ISM to SNR ejecta is simply $M_{SISM}/M_{EJ} = (4\pi/3)\, R_{TS}^3 \rho_o/M_{EJ}$, and $\rho_o$ is the ISM mass density. Since at the onset of the Sedov-Taylor expansion stage the characteristic mass ratio $M_{SISM}/M_{EJ} = 1.6$ (McKee & Truelove 1995), and the homologous relative shock radius $R*/R_{ST} \sim (t*/t_{ST})^{2/5}$, the age-dependent mixing mass ratio is simply $M_{SISM}/M_{EJ} \sim 1.6 \, (t*/t_{ST})^{6/5}$ during Sedov-Taylor expansion.

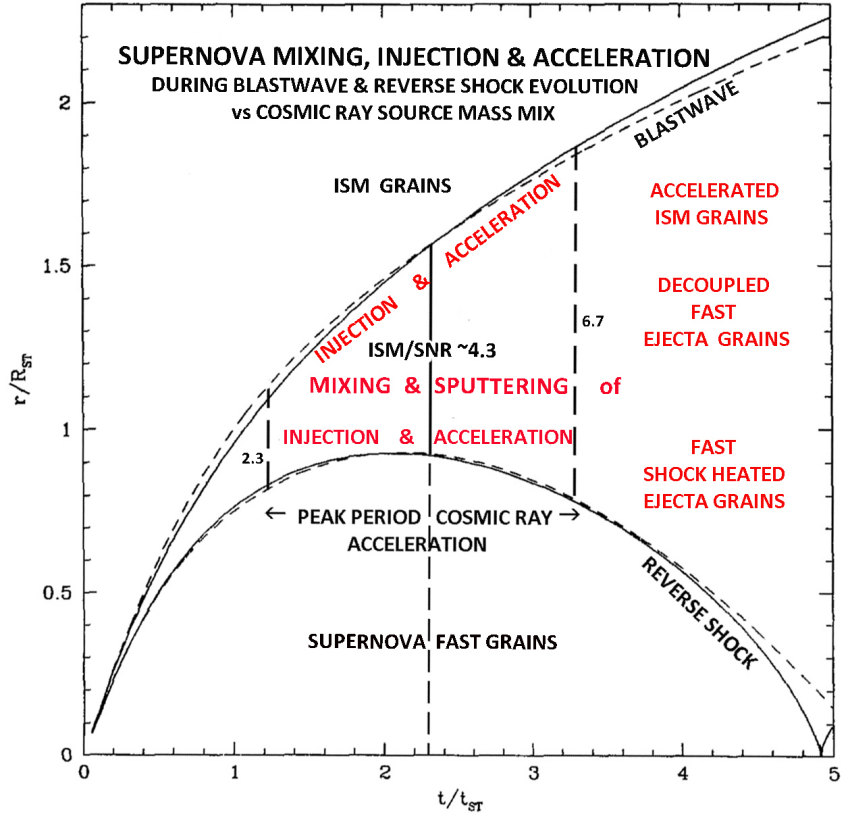

Figure 11. Homologous Blastwave and Reverse Shock Radii versus Age show that the measured cosmic ray mixing mass ratio of ISM/CCSN ~ 4.3 (Murphy et al. 2016) occurs right while the reverse shock radius and energy are passing through their peak, when the shock is decoupling fast massive, low-charge ejecta grains from slowing ejecta plasma. These grains move on outward to mix with swept-up and accelerated ISM grains, sputtering and injecting more suprathermal ions into diffusive shocks that accelerate cosmic rays at the current mix (shock curves adapted from McKee & Truelove 1995).

Thus, the measured mean cosmic ray mixing mass ratio of 4.3 (Murphy et al. 2016) is equal to the homologous mixing value calculated in supernova remnants around ~2.3 times the Sedov-Taylor onset time, $t*/t_{ST}$ on the reverse shock evolutionary track (McKee & Truelove 1995). For the principal $M_{EJ}$ ~10 to 20 $M_\odot$ CCSNe in superbubbles of ~0.001 H cm$^{-3}$ the peak acceleration occurs at ages of ~30 to 60 kyr (Figure 10). Also as noted before, even with the ISM dominating the mass by a factor of ~4:1, the still larger factor of ~10 metallicity of the ejecta still dominates ~2:1 by element over ISM.

The composition of the cosmic rays is once again the key that graphically ties together the ensemble of homologous, and concurrent processes of cosmic ray acceleration driven by supernova shock and turbulence generation. The reverse shock decouples the fast ejecta grains from the slowing ejecta plasma, which they leave behind and move forward to mix with the swept-up and accelerated ISM. In this growing turbulent mix between the forward and reverse shocks, these grains sputter off enriched suprathermal ions, injecting them into the diffusive shock acceleration that carries them to cosmic ray energies. Through numerical simulations of the blastwave and reverse shock trajectories versus time McKee & Truelove (1995) map out the homologous temporal and spatial structure within the expanding supernova remnants during that peak period of cosmic ray generation (Figures 10 & 11), around the mean cosmic ray mixing mass ratio of ~ 4.3, at ~2.3 times the Sedov-Taylor onset time $t*/t_{ST}$.

All of the shocked, swept-up ISM and CCSN ejecta at any instance lies in the growing shell of the remnant between the blastwave and reverse shocks, and between both shocks are the highly turbulent regions (e.g.Chevalier & Fransson 2003), where most grain-gas interactions that inject suprathermal ions into those shocks occur. Diffusing back and forth into these confining shocks, these fast ions are differentially accelerated to cosmic ray energies.

The peak cosmic ray acceleration occurs quite early, just ~5% , or ~20 kyr out of the full ~400 kyr Sedov-Taylor stage. This peak also occurs together with the strong shocks and hard compression ratios of $s$ ~ 3 to 4 that can accelerate cosmic rays to power law energy spectra of $\gamma = (s + 2)/(s -1)$ and are seen in cosmic ray produced $\pi^0$-decay $\gamma$'s from both supernovae in ISM (Acero e al. 2016) and in superbubbles (Ackermann et al. 2011; Aharonian et al 2018).

### 5.3  Dust Grain Survival in Supernova Ejecta

The role of the reverse shock, however, can shift from the driving factor in cosmic ray metal enrichment by Coulomb sputtering injection of suprathermal ions from fast grains in hot ~$10^6$ K, tenuous ~0.001 H cm$^{-3}$ superbubbles, to becoming the most destructive factor in stopping such grain enrichment in all of the cooler, denser >0.01 H cm$^{-3}$ phases of the ISM, where the sputtering rate is orders of magnitude higher, nearly completely destroying all grains.

As we saw, nearly all of the refractory elements condense into fast dust grains early in freely expanding ejecta, as the supernovae rapidly cool to <100 K in the moving frame. But with the development of a reverse shock from swept-up ISM during the onset of the Sedov-Taylor stage and its slowing of the ejecta plasma, grains are decoupled (Lingenfelter et al. 1998; Bianchi & Schneider 2007) from the shocked plasma and move on ahead at suprathermal velocities of nearly their free expansion speed of ~2,000 to 3,000 km s$^{-1}$ into the slowing ejecta plasma and swept-up, compressed and mixing ISM. There grain atoms interacting with the turbulent volatiles, are sputtered off the grain as suprathermal ions at nearly the same speed, >10 X that of ~$10^6$ K superbubble thermal H, or as much as an MeV per atom for major refractories, far above the sputtering surface binding energies of < 10 eV/atom (e.g. Robinson 1981) and injected into the diffusive shock acceleration process to be carried to cosmic ray energies with high efficiency in ~ 0.001 H cm$^{-3}$ superbubbles. But as noted, for supernovae occurring in the typical >0.01 H cm$^{-3}$ ISM on the other hand, the reverse shock generates much higher sputtering interaction rates at this same phase of the Sedov-Taylor transition that virtually break up the entire grains.

Here we consider specifically the survival of these grains in the light of their essential role in accelerating the heavy elements Z>2 in the cosmic rays. This distinctive signature of ejecta and accelerated ISM grain Coulomb sputtering in the $Z^{2/3}$-dependent (Sigmund 1969, 1981) abundance enrichment of cosmic ray metals, shown in Figure 5 above, clearly demands that in the major cosmic-ray accelerating sources the bulk of the fast ejecta grains escape destruction during the reverse shock passage through the cosmic ray metal source mix, which, as the mixing ratio requires, occurs at the same homologous time as the peak shock acceleration of the metals in supernova remnants. The observation of that Z-enrichment requires that only a fraction of the grains can be sputtered, since if the sputtering and grain-grain collisions completely broke up the grains, ionizing all of the elements, their relative abundances would then be determined just by their bulk composition, and there would be *no remaining differential enrichment*.

Detailed studies of grain survival have shown that even modest reverse shocks can still drive extensive sputtering and grain-grain collisions that destroy > 90% of the ejecta grains in supernovae that occur in the typical interstellar gas denser than $n > 0.1$ H cm$^{-3}$ (Bianchi & Schneider, 2007, Fig. 5; Micelotta et al. 2018; Sarangi et al. 2018). But, in all of the CCSN remnants expanding in the much lower density $n < 0.01$ H cm$^{-3}$ superbubble interiors grain sputtering is only partial and > 50% of the ejecta grains can survive the reverse shock, generating the observed Z –dependent abundance enrichment.

The issue of grain survival is also important for the question of acceleration of cosmic ray metals in SN Ia thermonuclear supernovae, which because of the much greater age of their progenitors would spread far beyond their natal OB association and be expected to produce the bulk of the single supernova remnants in the interstellar medium, where typical densities range from 0.1-1 H cm$^{-3}$ and grains are not expected to survive. The consistency of the Fe group cosmic ray abundances with those expected just from CCSN, and thus the lack of any sign of the large Fe component expected from the SN Ia, if they were accelerated with equal efficiency as CCSN, e.g. Section 4.2, also strongly argue for the effect of ambient density on grain survival and cosmic ray metal acceleration. This also raises the question of whether even in superbubbles grains survive in the SN Ib and Ic component of CCSN, since they are thought to arise (e.g. Branch & Wheeler 2017) from the most massive stars, >40 M$_\odot$ O stars after they have lost most of their H envelopes either to binary companions or through powerful early winds. So after exploding such supernova remnants would be expected to expand into denser, ~10$^6$ H cm$^{-3}$ (Chevalier & Fransson 2003), much more massive circumstellar remnants of their progenitor and companion winds, where grains would be very efficiently destroyed.

Thus the essential need for grain sputtered metal injection to explain the cosmic ray source composition strongly suggests that the dominant ~8 to ~25 M$_\odot$ SN II component of CCSN in tenuous superbubbles surrounding OB-association star formation regions are not only the primary energy source of Galactic cosmic ray protons and He, but virtually their sole effective source of heavy nuclei Z>2.

Thus our understanding of the key roles of mixing and injection in cosmic ray composition has grown greatly in the last couple decades, since Meyer et al. (1997) and Ellison et al. (1997) first introduced swept-up ISM grain acceleration in Sedov-Tayor expanding supernovae and their simultaneous grain sputtered suprathermal ion injection of refractories into quantitative models of diffusive shock acceleration of cosmic rays. Following the roughly similar (<2x) $(Z/H)_{GCRS/SS}$ ratios of most highly refractory cosmic ray abundances to solar system values, they assumed a roughly constant sputtering rate of ~ $0.01nva^2$ with a cross section based only on grain size *a*, appropriate to grain-grain collisions, which gives no measure of elemental abundances. They did not use the standard $Z^{2/3}$-dependent Coulomb sputtering cross section of Sigmund (1969;1981), which only clearly revealed itself (Rauch et al. 2009; Murphy et al. 2016; and Figure 5 above) after the ~4:1 bulk mass ISM/Ejecta mix was seen to be ubiquitous. Ellison et al.'s (1997) model, therefore, instead produces a net, charge- and mass-independent enhancement of refractories, while they made the accompanying cosmic ray volatile abundance enhancements proportional to atomic mass *A/Q* for an assumed weakly ionized *Q*~ 2 gas. These thus roughly matched with what were the available cosmic ray measurements (Englemann et al., 1990) and analyses at that time.

Ellison et al. (1997), however, also introduced supernova shock acceleration of a bulk mixing of both swept up ISM and circumstellar (CSM) gas and grains from earlier progenitor winds. Yet they did not include the ongoing mixing with the high velocity grains of the high metallicity supernova ejecta itself that was driving the turbulent early Sedov-Taylor expansion. As we have seen in Figure 11, this mixing was already modelled in detail (McKee & Truelove 1995). But unambiguous cosmic ray mixing data was not yet there. Recent measurements and analyses (Rauch et al. 2009; Ahn et al. 2010; Murphy et al. 2016) based on the ubiquitous ~4:1 bulk mixing ratio, now no longer suggest significantly different behavior between highly refractory and volatile cosmic ray abundances, and both effectively show atomic charge-dependent Coulomb enhancements proportional to $Z^{2/3}$ (Figure 5), while volatile implantation in grains has also become recognized.

## 6. SUMMARY AND FURTHER WORK

Far exceeding the expectations of Baade & Zwicky (1934a,b) and Ginzburg & Syrovatskii (1964), measurements of the abundances of the Galactic cosmic ray metals $(Z/H)_{GCRS}$ have defined a self-consistent framework of their sources, sites, and processes of mixing, injection and acceleration. Their elemental composition is basically that of core collapse supernova ejecta, mixed and partially diluted with swept-up interstellar medium, differentially enriched by grain condensation and sputtering injection into diffusive shock acceleration. Their composition has in fact defined these two basic processes so clearly that they permit the construction of the elemental cosmic ray abundances to within a $1\sigma$ uncertainty of ±35% with no free parameters. Moreover, the measured array of elemental cosmic ray source abundances determines the ubiquitous bulk mass mixing ratio of swept-up interstellar gas to CCSN ejecta to such a degree , ~ 4:1, that it identifies the time span within the homologous remnant expansion, when this mixing, injection, and acceleration occurs shortly after the onset of the decelerating, shock-generating, turbulent Sedov-Taylor stage of expansion and mixing.

Core collapse supernovae are also by far ~81%, the most frequent in our Galaxy and the major sources of both its supernova ejecta and shock accelerating power. The cosmic ray metals are not accelerated by all of the core collapse supernovae, however, only those that occur in the hot ~ $10^6$ K, low density 0.01-0.001 H $cm^{-3}$ superbubble gas. There the reverse shocks of remnants expanding into such low densities are not strong enough to destroy more than a small fraction of the crucial refractory grains necessary to produce the observed $Z^{2/3}$-dependent sputtering injection yields. Also in these low densities cosmic ray accelerating shocks enjoy much lower ambient ionization losses, and are far more efficient (e.g. Axford 1981).

Thus, the robust success of both the nonrotating and rotating models (Woosley & Heger 2007; Limongi & Chieffi 2018) of the CCSN ejecta compositions, as mixed with swept-up ISM, in clearly organizing and revealing the refractory grain injection processes (Section 3, Figures 5-8) provides strong evidence that CCSN are not just the primary source of Galactic supernova shock energy and cosmic ray proton and He, whose ratio still involves all supernovae, but nearly the sole source of cosmic ray metals, Z>2.

This is particularly reinforced by the fact that the cosmic ray composition measurements have shown no significant evidence of the dominant Fe-group contribution, expected (Section 4.2) from SNe Ia, based on their measured luminosity and frequency. These cosmic ray measurements do, nonetheless, strongly show additional C and N contributions, expected (Section 4.1) from Wolf-Rayet winds of some of the most massive (> 40 M$_\odot$) CCSN progenitors.

We have also seen that extensive astronomical observations of star forming regions and detailed measurements of the unusual cosmic ray composition have provided compelling evidence that at a minimum over 75% of the Galactic cosmic rays from about 0.1 GeV up to the "ankle" at around $10^9$ GeV are accelerated by supernova shocks out of highly ejecta-enriched dust and gas by spatially and temporally clustered bursts of core collapse supernovae, SNII and SNIb/c, in superbubbles formed around massive OB associations, while most of the remainder of the cosmic rays are accelerated by the shocks of single thermonuclear supernovae, SNIa, scattered randomly throughout the interstellar medium.

Not all is explained, of course, for the source of the very highest energy cosmic rays above the ankle at ~$10^9$ GeV is still unknown (e.g. Olinto 2013).

But these abundance measurements have explained the origin of by far the great bulk of the Galactic cosmic rays. For we see that applying these basic processes of mixing, condensation and injection to solar system abundances can thus match major cosmic ray abundances to within ±35% uncertainty with no free parameters.

Thus we've seen just how truly revealing the cosmic ray abundance ratios of the metals, Z>2, can be in probing critical nuclear, atomic and even solid state processes in addition to plasma processes, that are not accessible from the study of the H and He ratio alone, which is actually dominated by the interstellar medium unlike the metals. In particular, the metals give multiple measures of the mass mixing ratio of the ejecta and the ambient ISM in the cosmic ray acceleration region, together with the nucleosynthetic yields of the ejecta that can identify the type of supernova or other source. They also probe major atomic and mineralogical processes, including differential ionization and grain condensations fractions that determine the relative elemental ion injection factors that dominate both the cosmic ray abundances and their overall acceleration efficiencies. Thus the metals measure major processes not accessible from plasma studies alone.

There is obviously still much to be done. What we need to explore next is the impact and interplay of these cosmic ray compositional source constraints on the details of the diffusive shock acceleration of the cosmic rays. In particular, it would appear that simulations are most needed of diffusive shock acceleration by both blastwave and reverse shocks in both the swept-up ISM and the high metallicity, grain loaded, ejecta, respectively, of core collapse supernovae exploding in the hot ~$10^6$ K, tenuous ~0.001 H cm$^{-3}$, turbulent gas and magnetic fields of superbubbles surrounding OB association-star formation regions. These are the particular source conditions required by the cosmic ray compositional constraints, but they have been commonly overlooked in favor of the more common warm, denser phases of the ISM, which we now find are *not consistent* with the measured cosmic ray composition.

Such cosmic ray acceleration simulations need to be made in connection with homologous models of supernova remnant evolution, and i) grain freezeout fractionization during free-expansion at >2000 km s$^{-1}$, ii) early Sedov-Taylor mixing of ejecta and swept-up ISM that produces the ubiquitous cosmic ray mass ratio, iii) high speed grain sputtering with the classical $Z^{2/3}$ Coulomb cross section (e.g. Sigmund 1969) to produce suprathermal ions, and iv) injects enriched metals into the cosmic rays, that are dictated by both the measured cosmic ray composition and supernova hydrodynamics, rather than a simple grain-grain collision estimate, or by the commonly assumed partially ionized ISM alone injected by first-ionization potential, which are also *not at all consistent* with cosmic ray measurements. Such are the new simulations that can help put together a general integrated program to explore how these interact to generate the bulk of the local cosmic rays in measureable detail.

As we have also seen, the cosmic ray source composition is quite consistent with a mix of the swept-up ISM and the ejecta of the most common ~81%, supernovae, the CCSNe, or SN II & Ib/c. The bulk ~70% (Shivvers et al. 2017) of these are the SN II with ejecta masses of roughly ~10 $M_\odot$ to ~20 $M_\odot$ and Main Sequence lifetimes of ~12 to 35 Myr before they explode. But only a fraction <1/3 of the SN II come from the prominent O-stars all of which explode by the first ~13 Myr of an OB association's life, leaving all of the remaining SN II progenitors uncounted in the surveys of OB associations which are based only on the O-star counts. Recently, however, (Comeron et al. 2016) has successfully surveyed the early B-star Red supergiants in the Cygnus OB2 association significantly increasing the potential CCSN count in that association and since these are roughly half of the cosmic ray sources more such surveys need to be made.

Finally, in order to determine the sources of the explosive nucleosynthetic r-process, it is particularly important to measure the cosmic ray source abundances of the major r-process elements and particularly the actinides to higher precision, and this should be pursued with high priority.

**Acknowledgements**

I am particularly indebted to my old friend James C. Higdon of Claremont Colleges with whom I collaborated on much of this early work; to my close colleague Richard E. Rothschild of UCSD for many stimulating discussions, to W. Robert Binns of Washington University for very valuable comments on cosmic ray measurements and modeling, and especially to the anonymous reviewer's very thoughtful, incisive comments and criticisms that have greatly helped to focus and improve this paper.

# APPENDIX

As a handy reference, definitions of the major notations, used here in developing the transformation of the mix of the supernova ejecta and swept-up interstellar abundances into that of cosmic rays, are given in the Appendix in Table 1.

**Table 1  List of Notations**

| | |
|---|---|
| $C_{R/V}$ | cosmic ray refractory/volatile enrichment constant |
| $E_{51}$ | supernova blast energy, $10^{51}$ erg |
| $F_{CS}$ | Coulomb sputtering and scattering injection |
| $F_{EJ}$ | supernova ejecta bulk mixing mass fraction |
| $F_{GC}$ | grain condensation fraction |
| $F_{GI}$ | elemental grain injection factor |
| $F_{ISM}$ | swept-up ISM bulk mixing mass fraction |
| $F_V$ | volatile implantation fraction |
| $F_X$ | unknown source fractions |
| $M_{EJ}$ | Ejecta mass |
| $M_{SISM}$ | Swept-up ISM mass |
| $n_O$ | density, H cm$^{-3}$ |
| $R_{ST}$ | Sedov-Taylor radius |
| $t_{ST}$ | Sedov-Taylor onset time |
| $V_{FE}$ | Free expansion velocity, km s$^{-1}$ |
| $(Z/H)_{CCSN}$ | Core collapse supernova ejecta abundances |
| $(Z/H)_{EJ}$ | Supernova ejecta abundances |
| $(Z/H)_{GCRS}$ | Galactic cosmic ray source abundances |
| $(Z/H)_{ISM}$ | ISM abundances, ~1.32 $(Z/H)_{SS}$ |
| $(Z/H)_{SM}$ | source-mix model abundances |
| $(Z/H)_{SMCSI}$ | Coulomb sputtering-injected, source-mix model abundances |
| $(Z/H)_{SMGI}$ | grain-injected, source-mix model abundances |
| $(Z/H)_{SS}$ | Solar system abundances |
| $(Z/H)_{XMI}$ | Potential extra source abundances |

Quantitative values of the major steps along the way, shown in Figures 2 through 8, are also listed in the following Tables Appendix A and B.

# APPENDIX A  Non-Rotating Stars

| | Z | (Z/H)$_{GCRS}$[1] | (Z/H)$_{SS}$[2] | GCRS/SS Fig. 2 | (Z/H)$_{EJ/SS}$[3] | (Z/H)$_{EJ}$ | (Z/H)$_{ISM}$ | (Z/H)$_{SM}$ | SM/SS Fig. 4 | GCRS/SM Fig. 5 | F$_{CS}$ | GCRS/SMCSI Fig. 6 | F$_{GC}$[4] | GCRS/SMGI Fig. 8 |
|---|---|---|---|---|---|---|---|---|---|---|---|---|---|---|
| H | 1 | 10000 | 10000 | 1.00 | 1.00 | 10000 | 10000 | 10000 | 1.00 | 1.00 | 10000 | 1.00 | 1.00 | 1.00 |
| He | 2 | 4016 | 963 | 4.17 | 2.15 | 2068 | 963 | 1184 | 1.23 | 3.39 | 1468 | 2.74 | 1.00 | 2.74 |
| C | 6 | 166±5 | 2.91 | 57±2 | 28.0 | 81.6 | 3.78 | 19.3 | 6.63 | 8.6±0.3 | 55.8 | 2.97±0.09 | 1.39 | 2.14±0.06 |
| N | 7 | 14.9±0.8 | 0.80 | 19±1 | 6.02 | 4.81 | 1.04 | 1.79 | 2.24 | 8.3±0.5 | 5.80 | 2.57±0.14 | 1.14 | 2.25±0.12 |
| O | 8 | 208±4 | 5.81 | 36±1 | 30.8 | 179 | 7.55 | 41.8 | 7.19 | 5.0±0.1 | 149 | 1.40±0.03 | 1.66 | 0.84±0.02 |
| Ne | 10 | 23.6±0.3 | 0.88 | 27±2 | 28.4 | 25.0 | 1.14 | 5.91 | 6.72 | 4.0±0.05 | 24.8 | 0.95±0.01 | 1.00 | 0.95±0.01 |
| Na | 11 | 1.63±0.11 | 0.024 | 68±5 | 22.5 | 0.54 | 0.031 | 0.13 | 5.42 | 12.5±0.9 | 0.59 | 2.76±0.19 | 2.84 | 0.97±0.07 |
| Mg | 12 | 45.1±0.5 | 0.42 | 107±1 | 14.6 | 6.13 | 0.55 | 1.67 | 3.98 | 27.0±0.3 | 8.02 | 5.62±0.06 | 4.64 | 1.21±0.01 |
| Al | 13 | 3.93±0.34 | 0.035 | 112±10 | 15.3 | 0.54 | 0.046 | 0.14 | 4.00 | 28.1±2.4 | 0.71 | 5.54±0.48 | 5.50 | 1.00±0.08 |
| Si | 14 | 40.6±0.1 | 0.412 | 98.5±0.2 | 12.7 | 5.24 | 0.54 | 1.48 | 3.59 | 27.4±0.1 | 7.92 | 5.13±0.01 | 4.65 | 1.10±0.01 |
| P | 15 | 0.29±0.08 | 0.0034 | 85±24 | 16.9 | 0.057 | 0.0045 | 0.015 | 4.41 | 19±5 | 0.085 | 3.4±0.9 | 3.43 | 0.99±0.26 |
| S | 16 | 5.33±0.08 | 0.18 | 30±1 | 15.3 | 2.76 | 0.23 | 0.74 | 4.11 | 7.2±0.1 | 4.37 | 1.22±0.02 | 1.86 | 0.66±0.01 |
| Ar | 18 | 0.76±0.07 | 0.042 | 18±2 | 9.84 | 0.41 | 0.055 | 0.13 | 3.10 | 5.9±0.5 | 0.83 | 0.92±0.08 | 1.00 | 0.92±0.08 |
| Ca | 20 | 2.55±0.04 | 0.026 | 98±2 | 8.41 | 0.22 | 0.034 | 0.071 | 2.73 | 35.9±0.6 | 0.49 | 5.20±0.08 | 5.60 | 0.93±0.01 |
| Fe | 26 | 45.2±1.5 | 0.345 | 131±4.4 | 7.93 | 2.74 | 0.449 | 0.91 | 2.64 | 49.7±1.7 | 7.53 | 6.00±0.20 | 4.22 | 1.42±0.05 |
| Co | 27 | 0.10±0.04 | 0.00096 | 104±42 | 14.3 | 0.014 | 0.0013 | 0.0038 | 4.00 | 26.3±10.5 | 0.032 | 3.13±1.25 | 4.21 | 0.74±0.30 |
| Ni | 28 | 2.44±0.04 | 0.020 | 122±2 | 9.77 | 0.20 | 0.026 | 0.061 | 3.05 | 40.0±0.7 | 0.53 | 4.60±0.08 | 4.07 | 1.13±0.02 |
| Cu | 29 | 225±37 | 2.17 | 104±17 | 41.0 | 88.9 | 2.86 | 20.1 | 9.26 | 11.2±1.8 | 180 | 1.25±0.21 | 3.26 | 0.38±0.06 |
| Zn | 30 | 314±19 | 5.05 | 62.2±3.8 | 14.6 | 73.7 | 6.57 | 20.0 | 3.96 | 15.7±1.0 | 184 | 1.71±0.10 | 1.84 | 0.93±0.05 |
| Ga | 31 | 27±3 | 0.148 | 182±20 | 49.2 | 7.28 | 0.19 | 1.61 | 10.9 | 16.8±1.9 | 15.1 | 1.78±0.20 | 2.62 | 0.68±0.08 |
| Ge | 32 | 40±4 | 0.496 | 81±8 | 30.3 | 15.0 | 0.64 | 3.51 | 7.08 | 11.4±1.1 | 33.7 | 1.19±0.12 | 2.44 | 0.49±0.05 |
| As | 33 | 5.5±1.5 | 0.025 | 220±60 | 36.6 | 0.92 | 0.033 | 0.21 | 8.40 | 26.2±7.1 | 2.06 | 2.67±0.73 | 3.34 | 0.80±0.22 |
| Se | 34 | 17.6±2.4 | 0.271 | 65±9 | 21.8 | 5.91 | 0.35 | 1.46 | 5.39 | 12.1±1.6 | 14.6 | 1.20±0.16 | 1.99 | 0.60±0.08 |
| Br | 35 | 5.0±1.4 | 0.047 | 106±30 | 28.4 | 1.3 | 0.061 | 0.31 | 6.60 | 16.1±4.5 | 3.17 | 1.58±0.44 | 1.95 | 0.81±0.23 |
| Kr | 36 | 7.8±1.4 | 0.227 | 34±6 | 14.9 | 3.38 | 0.30 | 0.92 | 4.05 | 8.5±1.5 | 9.59 | 0.81±0.15 | 1.00 | 0.81±0.15 |
| Rb | 37 | 5.3±1.4 | 0.027 | 196±52 | 30.3 | 0.82 | 0.035 | 0.19 | 7.04 | 27.9±7.4 | 2.02 | 2.62±0.69 | 2.95 | 0.89±0.24 |
| Sr | 38 | 15.2±2.1 | 0.097 | 157±22 | 8.4 | 0.81 | 0.13 | 0.27 | 2.78 | 56.3±7.8 | 2.92 | 5.20±0.72 | 5.30 | 0.98±0.14 |
| Zr | 40 | 6.2±1.3 | 0.047 | 132±28 | 3.9 | 0.18 | 0.06 | 0.085 | 1.81 | 73±15 | 1.00 | 6.5±1.4 | 6.15 | 1.06±0.23 |
| Ba | 56 | 2.1±0.3 | 0.0179 | 117±17 | 2.4 | 0.043 | 0.023 | 0.027 | 1.51 | 78±11 | 0.40 | 5.3±0.8 | 5.86 | 0.90±0.14 |
| Hf | 72 | 0.14± | 0.00070 | 200± | 1.85 | 0.0013 | 0.00091 | 0.0010 | 1.43 | 140± | 0.017 | 8.1± | 4.77 | 1.70± |
| W | 74 | 0.11± | 0.00053 | 208± | 1.87 | 0.0010 | 0.00069 | 0.00075 | 1.42 | 147± | 0.013 | 8.3± | 5.26 | 1.58± |
| Pt | 78 | 0.97±0.10 | 0.0056 | 173±18 | 1.41 | 0.0079 | 0.0073 | 0.0074 | 1.32 | 131±14 | 0.14 | 7.2±0.7 | 4.01 | 1.80±0.17s |
| Pb | 82 | 0.42±0.08 | 0.0134 | 31±6 | 2.00 | 0.027 | 0.0174 | 0.0193 | 1.44 | 22±4 | 0.36 | 1.15±0.23 | 1.00 | 1.15±0.23 |

Refs. 1 Engelmann et al. 1990, Cummings et al 2016 (Mean 1<Z<28), Rauch et al. 2009, Murphy et al 2012 (30<Z<40), Binns et al 1989 (40 < Z <70), Donnelly et al 2012 (Z > 70); Lodders 2003; 3 Woosley & Heger 2007, Fig.4;  4 Wasson & Kallemeyn 1988  (C3/C1), normalized to 1.00 for mean of Al, Ca, Sr, Ba & W

## APPENDIX B Rotating Stars CCSN

| | Z | $(Z/H)_{GCRS}$[1] | $(Z/H)_{SS}$[2] | GCRS/SS Fig. 2 | $(Z/H)_{EJ/SS}$[3] | $(Z/H)_{EJ}$ | $(Z/H)_{SM}$ | $(Z/H)_{SM}$ | SM/SS Fig. 4 | CRS/SM Fig. 5 | $F_{CS}$ | CRS/SMCSI Fig. 6 | $F_{GC}$[4] | CRS/SMGI Fig. 8 |
|---|---|---|---|---|---|---|---|---|---|---|---|---|---|---|
| H  | 1  | 10000     | 10000  | 1.00      | 1.00 | 10000 | 10000  | 10000 | 1.00 | 1.00      | 10000 | 1.00       | 1.00 | 1.00       |
| He | 2  | 4016      | 963    | 4.17      | 2.12 | 2038  | 963    | 1178  | 1.22 | 3.41      | 1465  | 2.73       | 1.00 | 2.73       |
| C  | 6  | 166±5     | 2.91   | 57±2      | 20.4 | 50.3  | 3.78   | 14.9  | 5.11 | 11.2±0.3  | 43.1  | 3.86±0.12  | 1.39 | 2.78±0.08  |
| N  | 7  | 14.9±0.8  | 0.80   | 19±1      | 6.06 | 4.95  | 1.04   | 1.80  | 2.25 | 8.3±0.5   | 5.83  | 2.55±0.14  | 1.14 | 2.24±0.12  |
| O  | 8  | 208±4     | 5.81   | 36±1      | 23.6 | 137   | 7.55   | 33.4  | 5.76 | 6.2±0.1   | 120   | 1.73±0.03  | 1.66 | 1.04±0.02  |
| Ne | 10 | 23.6±0.3  | 0.88   | 27±2      | 19.8 | 17.4  | 1.14   | 4.40  | 5.00 | 5.37±0.07 | 18.5  | 1.28±0.02  | 1.00 | 1.28±0.02  |
| Na | 11 | 1.63±0.11 | 0.024  | 68±5      | 11.9 | 0.29  | 0.031  | 0.082 | 3.41 | 19.9±1.4  | 0.37  | 4.38±0.31  | 2.84 | 1.54±0.10  |
| Mg | 12 | 45.1±0.5  | 0.42   | 107±1     | 10.3 | 4.33  | 0.55   | 1.31  | 3.12 | 34.5±0.4  | 6.29  | 7.17±0.08  | 4.64 | 1.55±0.01  |
| Al | 13 | 3.93±0.34 | 0.035  | 112±10    | 10.4 | 0.36  | 0.046  | 0.11  | 3.13 | 35.9±3.1  | 0.56  | 7.05±0.61  | 5.50 | 1.28±0.11  |
| Si | 14 | 40.6±0.1  | 0.412  | 98.5±0.2  | 18.8 | 7.75  | 0.54   | 1.98  | 4.81 | 20.5±0.1  | 10.62 | 3.82±0.01  | 4.65 | 0.82±0.01  |
| P  | 15 | 0.29±0.08 | 0.0034 | 85±24     | 17.0 | 0.058 | 0.0045 | 0.015 | 4.41 | 19±5      | 0.085 | 3.4±0.9    | 3.43 | 0.99±0.26  |
| S  | 16 | 5.33±0.08 | 0.18   | 30±1      | 13.2 | 2.38  | 0.23   | 0.66  | 3.66 | 8.1±0.2   | 3.89  | 1.37±0.02  | 1.86 | 0.74±0.01  |
| Ar | 18 | 0.76±0.07 | 0.042  | 18±2      | 9.58 | 0.40  | 0.055  | 0.12  | 3.13 | 6.1±0.5   | 0.80  | 0.95±0.08  | 1.00 | 0.95±0.08  |
| Ca | 20 | 2.55±0.04 | 0.026  | 98±2      | 9.11 | 0.24  | 0.034  | 0.074 | 2.84 | 34.5±0.6  | 0.51  | 5.00±0.08  | 5.60 | 0.89±0.01  |
| Fe | 26 | 45.2±1.5  | 0.345  | 131.0±4.4 | 7.42 | 2.56  | 0.449  | 0.87  | 2.53 | 51.8±1.7  | 7.24  | 6.25±0.21  | 4.22 | 1.48±0.05  |
| Co | 27 | 0.10±0.04 | 0.00096| 104±42    | 10.9 | 0.0105| 0.0013 | 0.0031| 3.27 | 31.9±12.8 | 0.027 | 3.73±1.49  | 4.21 | 0.90±0.36  |
| Ni | 28 | 2.44±0.04 | 0.020  | 122±2     | 13.6 | 0.27  | 0.026  | 0.075 | 3.76 | 32.5±0.5  | 0.66  | 3.72±0.06  | 4.07 | 0.91±0.02  |
| Cu | 29 | 225±37    | 2.17   | 104±17    | 17.0 | 36.9  | 2.86   | 9.67  | 4.45 | 23.3±3.8  | 86.7  | 2.59±0.43  | 3.26 | 0.79±0.13  |
| Zn | 30 | 314±19    | 5.05   | 62.2±3.8  | 32.8 | 166   | 6.57   | 38.4  | 7.61 | 8.16±0.5  | 125   | 2.50±0.15  | 1.84 | 1.36±0.06  |
| Ga | 31 | 27±3      | 0.148  | 182±20    | 45.2 | 6.69  | 0.19   | 1.49  | 10.1 | 18.1±1.9  | 14.0  | 1.93±0.20  | 2.62 | 0.74±0.07  |
| Ge | 32 | 40±4      | 0.496  | 81±8      | 27.3 | 13.5  | 0.64   | 3.22  | 6.49 | 12.4±1.2  | 30.9  | 1.29±0.13  | 2.44 | 0.53±0.05  |
| As | 33 | 5.5±1.5   | 0.025  | 220±60    | 39.2 | 0.98  | 0.033  | 0.22  | 8.90 | 24.7±7    | 2.09  | 2.63±0.75  | 3.34 | 0.79±0.22  |
| Se | 34 | 17.6±2.4  | 0.271  | 65±9      | 17.6 | 4.78  | 0.35   | 1.24  | 4.56 | 14.3±2.0  | 12.4  | 1.42±0.20  | 1.99 | 0.71±0.10  |
| Br | 35 | 5.0±1.4   | 0.047  | 106±30    | 7.9  | 0.37  | 0.061  | 0.12  | 2.63 | 40.5±11.3 | 1.23  | 4.07±1.1   | 1.95 | 2.09±0.59  |
| Kr | 36 | 7.8±1.4   | 0.227  | 34±6      | 15.2 | 3.45  | 0.30   | 0.93  | 4.10 | 8.4±1.5   | 9.69  | 0.80±0.15  | 1.00 | 0.80±0.15  |
| Rb | 37 | 5.3±1.4   | 0.027  | 196±52    | 17.7 | 0.48  | 0.035  | 0.12  | 4.58 | 42±12     | 1.32  | 4.02±1.06  | 2.95 | 1.36±0.36  |
| Sr | 38 | 15.2±2.1  | 0.097  | 157±22    | 13.5 | 1.30  | 0.13   | 0.36  | 3.75 | 42.2±5.8  | 4.02  | 3.78±0.56  | 5.30 | 0.71±0.10  |
| Zr | 40 | 6.2±1.3   | 0.047  | 132±28    | 4.11 | 0.19  | 0.06   | 0.087 | 1.89 | 71.6±15.0 | 0.98  | 6.4±1.3    | 6.15 | 1.03±0.22  |
| Ba | 56 | 2.1±0.3   | 0.0179 | 117±17    | 1.9  | 0.034 | 0.023  | 0.025 | 1.40 | 84±12     | 0.35  | 5.9±0.8    | 5.86 | 1.01±0.14  |
| Pb | 82 | 0.42±0.08 | 0.0134 | 31±6      | 0.72 | 0.027 | 0.0096 | 0.016 | 1.19 | 26±5      | 0.29  | 1.43±0.28  | 1.00 | 1.43±0.28  |

Refs. 1 Engelmann et al. 1990/Cummings et al 2016 (Mean 1<Z<28); Rauch et al. 2009, Murphy et al 2012 (30<Z<40); Binns et al 1989 (40 < Z < 70). Donnelly et al 2012 (Z > 70); 2 Lodders 2003; 3 Limongi & Chieffi 2018 (CCSN See text); 4 Wasson & Kallemeyn 1988 (C3/C1), normalized to 1.00 for mean of Al, Ca, Sr, Ba & W